\documentclass[10pt,journal,compsoc]{IEEEtran}
\usepackage{graphicx}
\usepackage{booktabs}
\usepackage{amsmath}
\usepackage{amsfonts}
\usepackage{amssymb}
\usepackage{xcolor}
\usepackage{multirow}

\usepackage{algorithm,algpseudocode}

\ifCLASSOPTIONcompsoc
  \usepackage[nocompress]{cite}
  \usepackage[caption=false,font=normalsize,labelfont=sf,textfont=sf]{subfig}
\else
  \usepackage{cite}
  \usepackage[caption=false,font=footnotesize]{subfig}
\fi

\ifCLASSINFOpdf
  
\else
  
\fi

\newtheorem{definition}{Definition}

\begin{document}

\title{ConTIG: Continuous Representation Learning on Temporal Interaction Graphs}

\author{
        Xu~Yan,
        Xiaoliang~Fan,~\IEEEmembership{Senior~Member,~IEEE,}
        Peizhen~Yang,
        Zonghan~Wu,
        Shirui~Pan,~\IEEEmembership{Member,~IEEE,}
        Longbiao~Chen,
        Yu~Zang,
        and~Cheng~Wang,~\IEEEmembership{Senior~Member,~IEEE}
\IEEEcompsocitemizethanks{
\IEEEcompsocthanksitem Xu Yan, Xiaoliang Fan (the corresponding author), Peizhen Yang, Longbiao Chen, Yu Zang and Cheng Wang are with Fujian Key Laboratory of Sensing and Computing for Smart Cities, School of Informatics, Computer Science \& Technology Department, Xiamen University, Xiamen, China. \protect\\
E-mail: yanxu97@stu.xmu.edu.cn, fanxiaoliang@xmu.edu.cn, yangpz@stu.xmu.edu.cn, longbiaochen@xmu.edu.cn, zangyu7@126.com, cwang@xmu.edu.cn
\IEEEcompsocthanksitem Zonghan Wu is with Centre for Artificial Intelligence, FEIT, University of Technology Sydney, Australia.\protect\\
E-mail: zonghan.wu-3@student.uts.edu.au
\IEEEcompsocthanksitem Shirui Pan is with Department of Data Science and AI, Faculty of Information Technology, Monash University, Australia.\protect\\
E-mail: shirui.pan@monash.edu}
        }

%
%

\markboth{Journal of \LaTeX\ Class Files,~Vol.~14, No.~8, July~2021}%
{Shell \MakeLowercase{\textit{et al.}}: Bare Demo of IEEEtran.cls for Computer Society Journals}
%



\IEEEtitleabstractindextext{
\begin{abstract}

Representation learning on temporal interaction graphs (TIG) is to model complex networks with the dynamic evolution of interactions arising in a broad spectrum of problems. Existing dynamic embedding methods on TIG discretely update node embeddings merely when an interaction occurs. They fail to capture the continuous dynamic evolution of  embedding trajectories of nodes. In this paper, we propose a two-module framework named ConTIG, a continuous representation method that captures the continuous dynamic evolution of node embedding trajectories.  With two essential modules, our model exploit three-fold factors in dynamic networks which include \textit{latest interaction}, \textit{neighbor features} and \textit{inherent characteristics}. In the first update module, we employ a continuous inference block to learn the nodes’ state trajectories by learning from time-adjacent interaction patterns between node pairs using ordinary differential equations. In the second transform module, we introduce a self-attention mechanism to predict future node embeddings by aggregating historical temporal interaction information. Experiments results demonstrate the superiority of ConTIG on temporal link prediction, temporal node recommendation and dynamic node classification tasks compared with a range of state-of-the-art baselines, especially for long-interval interactions prediction.

\end{abstract}

\begin{IEEEkeywords}
Graph Embedding, Temporal Interaction Graph, Neural Ordinary Differential Equations, Graph Neural Networks.
\end{IEEEkeywords}}

\maketitle

\IEEEdisplaynontitleabstractindextext

\IEEEpeerreviewmaketitle

\IEEEraisesectionheading{\section{Introduction}\label{sec:introduction}}

\IEEEPARstart{G}{raph} representation learning has attracted a surge of research attention owing to the widespread existence of graph-structured data in the real world such as social networks, citation networks, and other user-item interaction systems \cite{survey_grl_arxiv_2017}. Learning graph embedding is a powerful approach of graph representation learning, which maps the characteristics of nodes to a low-dimensional vector space so that the proximities of nodes in topological space can be well reserved \cite{survey_ne_tkde_2018}. It has shown a great success in graph representation learning from shallow graph embedding methods \cite{deepwalk_kdd_2014,line_www_2015,node2vec_kdd_2016} to deep graph neural networks (GNNs) \cite{survey_tnnls_2020,gcn_iclr_2017,graphsage_nips_2017,gat_iclr_2018}.

Recently, representation learning on dynamic graph has attracted many research attention \cite{survey_2020_jmlr,survey_2020_arxiv, survey_arxiv_2021}, and they mainly model temporal graphs either as a sequence of snapshots \cite{burstgraph_2019_ijcai,evolvegcn_2020_aaai,grade_2020_arxiv,ctgcn_tkde_2020,dtine_tkde_2020,jiao2021temporal}, or as real events with timestamps \cite{tdgnn_www_2020,gtea_2020_arxiv,ctdne_www_2018,ties_2020_kdd,tgat_iclr_2020,tgn_icmlgrl_2020}. 

\begin{figure}[!t]
\centering 
\includegraphics[width=0.5\textwidth]{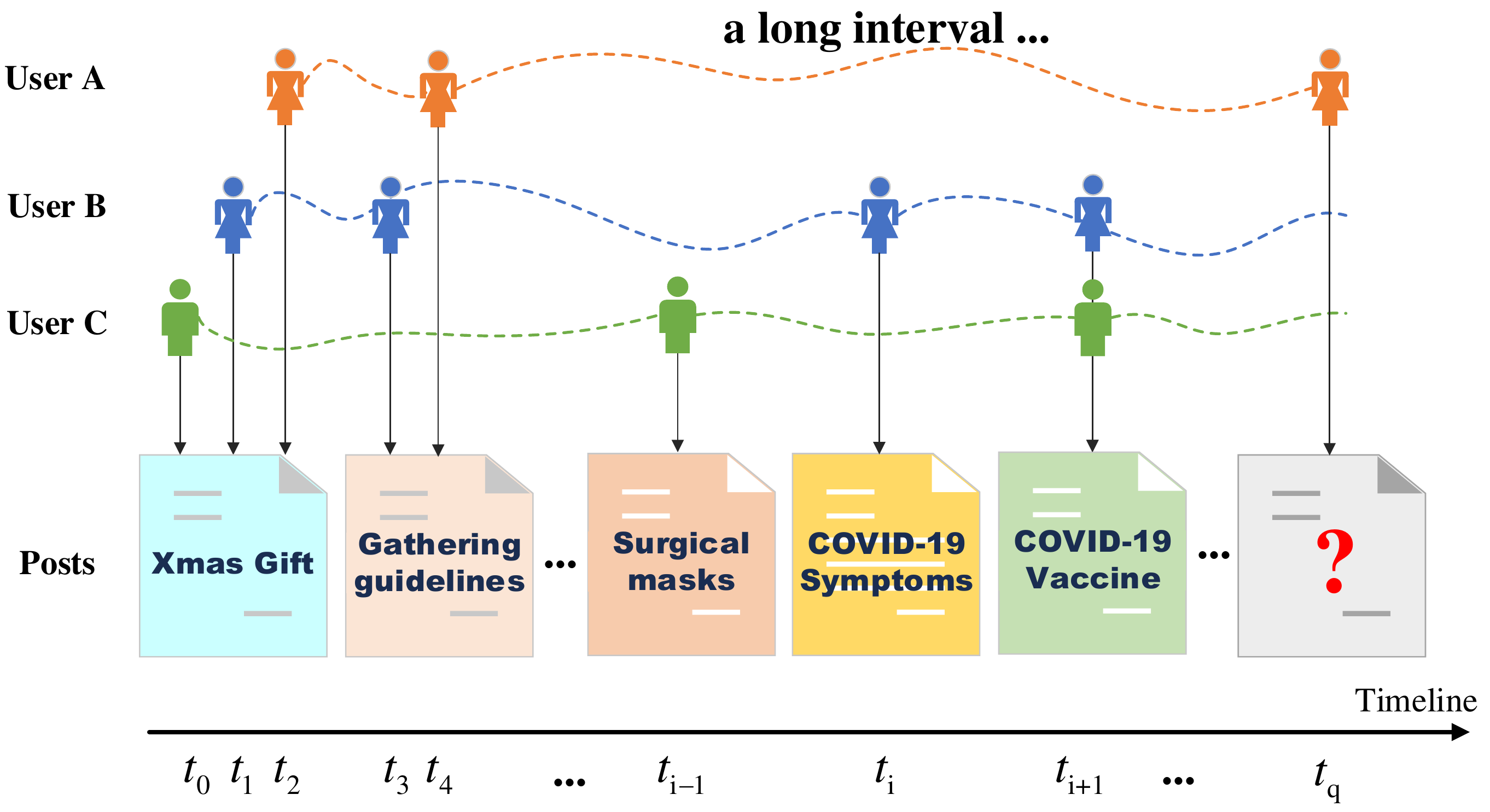}
\caption{An illustrated example of the inactive user (i.e., User A) with long-interval interaction in a temporal interaction graph. In the long-interval of User A, her future posts about COVID-19 might be affected by three factors: 1) her latest interaction with ``Gathering guidelines''; 2) her neighbors, user B and C; and 3) her inherent characteristics as a healthcare personnel.} 
\label{Fig.1}
\end{figure}

For existing works learning temporal interaction embedding from the sequence of interactions, a number of studies discretely update the embeddings of the interactive nodes once an interaction occurs \cite{jodie_kdd_2019,tigecmn_www_2020}. Other works learn the interactions by aggregating the temporal neighbor features to pass messages between nodes \cite{tdgnn_www_2020,tgat_iclr_2020}, which model discrete dynamic of node representation in multiple propagation layers \cite{discrete_dynamic_iclr_2020}. Although these discrete works have made substantial advances in learning temporal interaction embedding, they fail to capture the continuous dynamic evolution of nodes embedding trajectories thus losing temporal information for non-ignorable and inactive nodes with long-interval interactions (e.g., User A in Fig.\ref{Fig.1}). Therefore, we aim to learn nodes embedding trajectories and capture the continuous dynamic of node representations.

Learning dynamic node embedding trajectories is extremely challenging due to the complex non-linear dynamic in temporal interaction graphs (TIG). We conclude the challenges in three folds under the scenario of posting in Reddit\footnote{https://www.reddit.com/} website. \textbf{First}, \textit{latest interaction} information may have a significant impact on the current interaction, e.g., in Fig.\ref{Fig.1}, the posts made by User A at $t_4$ would be influential for $t_q$, since guidelines of holiday gathering under COVID-19 pandemics will be regularly implemented. \textbf{Second}, the node state is also affected by their \textit{neighbors' attention} over time. For instance, User A is expected to emulate their neighbors by posting ``COVID-19 vaccine'' at $t_q$ in Fig.\ref{Fig.1}. \textbf{Third}, \textit{inherent characteristics} of node would fatally determine the state regardless of aforementioned two factors. For example, User A is a frontline healthcare personnel and she could inherently pay much attention to COVID-19 trainings on infection control. In summary, existing methods hold partial considerations for these three-fold factors.

To cover the shortcomings of previous methods, we propose a \textbf{Con}tinuous representation learning method on \textbf{T}emporal \textbf{I}nteraction \textbf{G}raphs (\textbf{ConTIG}). The proposed ConTIG contains two modules: the update module and transform module. When an interactive message comes, in the update module, inspired by \cite{cgnn_icml_2020}, we define a neural ordinary differential equations (ODE)\cite{neuralode_nips_2018} with the three aforementioned factors affecting node states (i.e., latest interaction, neighbors' feature and inherent characteristics), and incorporate them with the continuous-time encoding function to trace dynamic knowledge and learn node state trajectories. Then the node embeddings at the ending time of neural ODE is used to update embeddings of interacting nodes. In the transform module, a self-attention mechanism is introduced to awaken historical interactive information of current interacting nodes and convert them to generate future node representations. The results on temporal link prediction, temporal node recommendation and dynamic node classification tasks show the effectiveness of our proposed model compared with a range of state-of-the-art baselines, especially for long-interval interactions prediction. 

The contributions of this work are summarized as follows:
\begin{itemize}
    \item We propose a representation learning method to incorporate three-fold factors (i.e., latest interaction, neighbor features and inherent characteristics) with a neural ODE. This allows us to capture continuous dynamic of node embedding trajectories by continuously updating node embeddings in the update module.
    \item We combine aforementioned update module with a self-attention mechanism to predict future embeddings in the transform module. This makes ConTIG effective on long-interval interactions prediction. 
    \item We evaluate our ConTIG on temporal link prediction, temporal node recommendation and dynamic node classification tasks, and ConTIG compares favorably against state-of-the-art methods.
\end{itemize}

The rest of this paper is organized as follows. In Section 2, we discuss some related work. Sections 3 and 4 describes the notations and our proposed model in detail. In Section 5, we conduct experiments on several datasets and compare them with state-of-the-art methods. In Section 6, the conclusion and our future work are presented.

\section{Related Work}
\subsection{Static Graph Embedding}

Early works for representation learning are mainly shallow models including graph factorization approaches \cite{le_nips_2001,gf_www_2013} and skip-gram models \cite{deepwalk_kdd_2014,line_www_2015,node2vec_kdd_2016}, which learn node representations by random walk objectives. With the success of deep learning, GNNs \cite{graphsage_nips_2017,gcn_iclr_2017,gat_iclr_2018,pan2019learning} have gradually attracted great research interest. They are effective approaches to learn node representations by updating each node according to messages from neighboring nodes in graphs in each layer. GNNs essentially model discrete dynamics of node representations with multiple propagation layers \cite{discrete_dynamic_iclr_2020}. However, all the above mentioned approaches are limited to learning node embeddings on static graph structure information, and the temporal behavior of interaction over time is generally ignored.

\subsection{Temporal Graph Embedding}
One common way to handle temporal graphs is to decompose it into multiple static graphs snapshots by a regular time interval. Some works embed the graph convolution into the recurrent neural network (RNN) based models or attention mechanism \cite{attention_nips_2017}, which learns to exploit the dynamic information in the graph evolution within a period of time \cite{burstgraph_2019_ijcai,tNodeEmbed_2019_ijcai,dysat_2020_icdm,evolvegcn_2020_aaai,ctgcn_tkde_2020,dtine_tkde_2020,htgn_kdd_2021}. Some other works are dynamic extensions of ideas applied in the static case inspired by methods such as PageRank\cite{dynamicppe_kdd_2021}, graph autoencoder\cite{dyngem_2018_arxiv} and the topic model \cite{grade_2020_arxiv} to capture both the temporal community dynamics and evolution of graph structure. However, learning embedding from the sequence of graph snapshots sampled from the temporal graph by a regular time interval may lose information by looking only at some snapshots of the graph over time. 

Therefore, recent works learn temporal graph embedding from the sequence of timed interactions. A number of studies learn the sequence of interactions by discretely updating the node embeddings once an interaction occurs by RNNs \cite{jodie_kdd_2019,trrn_aaai_2021,tgn_icmlgrl_2020}, memory networks \cite{tigecmn_www_2020}, time point process \cite{dyrep_2019_iclr,m2dne_cikm_2019}, transformer network\cite{tcl_arxiv_2021}, generative models \cite{taggen_kdd_2020}, etc. Other works learn the interactions by aggregating the temporal neighbor features to pass messages between nodes \cite{gtea_2020_arxiv,tgat_iclr_2020,ad_arxiv_2021,tgn_icmlgrl_2020} or learning node representation from random walk objects \cite{ctdne_www_2018} and temporal motifs \cite{motifs_wsdm_2017,caw_iclr_2021,lmega_kdd_2020}. TGAT \cite{tgat_iclr_2020} proposes the temporal graph attention layer to aggregate temporal-topological neighborhood features. TGN \cite{tgn_icmlgrl_2020} makes a combination of updating operators and aggregating operators. However, these methods learn the discrete dynamic of node representations, and it is not beneficial to learn the node state trajectories. Moreover, they ignore the influence of the inherent characteristics of the change on the node states, which is not beneficial to capture the complex non-linear dynamic of node representations. Different from the aforementioned works, we assembled the latest interaction, neighbor features and inherent characteristics of the nodes into a neural ODE to learn node state trajectories and update node embeddings. As a result, our method captures the continuous dynamic of node representation.

\subsection{Neural ODE}
Neural ordinary differential equations (ODE)\cite{neuralode_nips_2018} is an approach for modelling a continuous dynamics on hidden representation, where the dynamic is characterised through an ODE
parameterised by a neural network (i.e. ODE Solver). Recently, there are some works devote to use neural ODE \cite{neuralode_nips_2018} combined with GNN to characterise the continuous dynamics of node representations \cite{cgnn_icml_2020,ndcn_kdd_2020}. However, these methods are not built for temporal graphs, which are continuous-time GNNs learning the dynamic of node representation in static graph scenario to build "deeper" networks. For temporal graphs, \cite{tkgode_arxiv_2021} learns the evolution of the temporal knowledge graph dynamics over time using ODEs as a tool. CG-ODE \cite{cgode_kdd_2021} proposed a latent ODE generative model that learns the coupled dynamics of nodes and edges with a GNN based ODE in a continuous manner. Different from them (i.e., \cite{tkgode_arxiv_2021,cgode_kdd_2021}), we stand at the perspective of node state modeling in temporal interaction graph, using a differential equation integrating the three-fold factors (i.e., latest interaction, neighbor features and inherent characteristics). This allows us to successfully capture continuous dynamic of node embedding trajectories.

\section{Problem Definition} \label{Sec 3.1}
We summarize some notations and state the problem we want to solve in this section.

\begin{definition}[Temporal interaction graph]
Temporal interaction graph is a graph with temporal and attributed interaction edges between nodes. Temporal interaction graph is defined as a pair $\mathcal{G} = (\mathcal{V}, \mathcal{E} ) $, where $\mathcal{V}$ represents vertices sets, and $\mathcal{E} $ is a set of the sequences of interactions with time label between two vertices. An interaction $e$ is a tuple of form $(u, i, t, f^{ui})$, where $u$ and $i \in \mathcal{V}$ represent two vertices have an interaction at timestamp $t$, and $f^{ui}$ represents the feature of interaction $e$. 
\end{definition}

\begin{definition}[Temporal interaction graph embedding]
Given the temporal interaction graph $\mathcal{G}=(\mathcal{V}, \mathcal{E})$, we aim to learn a mapping function: $\phi : \mathcal V \rightarrow \mathbb{R}^{d}$ to map node features into a low-dimensional latent space and generate node embeddings $H \in \mathbb{R}^{ \left |\mathcal{V} \right | \times d} $ to represent nodes, where $d \ll \left |\mathcal{V} \right |$ representing the dimension of node embeddings.
\end{definition}

\begin{definition}[Interaction intervals]
In the temporal interaction graph, each interaction is an edge connected by a source node and a target node, so we can use source nodes as the label of interaction to calculate the time interval $\Delta t_e $ between current interaction and its latest interaction, where the latest interaction is the edge where the source node last appeared. For example, the latest interaction of the current interaction, $e_p^j = (u^j, i, t_p, f^{u i})$, is the interaction $e_q^{j-1}=(u^{j-1}, r_u, \tau_u, f^{u r_u})$ in which the node $u$ at the last time appeared. As a result, the time interval $\Delta t_{e_p}$ between $e_p^j$ and $e_q^{j-1}$ is $t_p-\tau_u$, where $r_u$ and $\tau_u$ represents the latest interactive node and timestamp of $u$, respectively. While, $p$ and $q$ represent the interaction id, and $j$ represents that how many times $u$ appears. Here, if the node appears for the first time (i.e., $j=0$), the time interval of the interaction is $0$.
\end{definition}

\section{The Proposed Model}
In this section, we will describe our model in detail. The proposed model – ConTIG consists of two essential modules (i.e., update and transform module). We will briefly introduce them in Section \ref{sec4.1} and describe each module in detail in the rest of sections.

\subsection{Overview of the ConTIG}\label{sec4.1}

\begin{figure*}[!t]
\centering
\includegraphics[width=\textwidth]{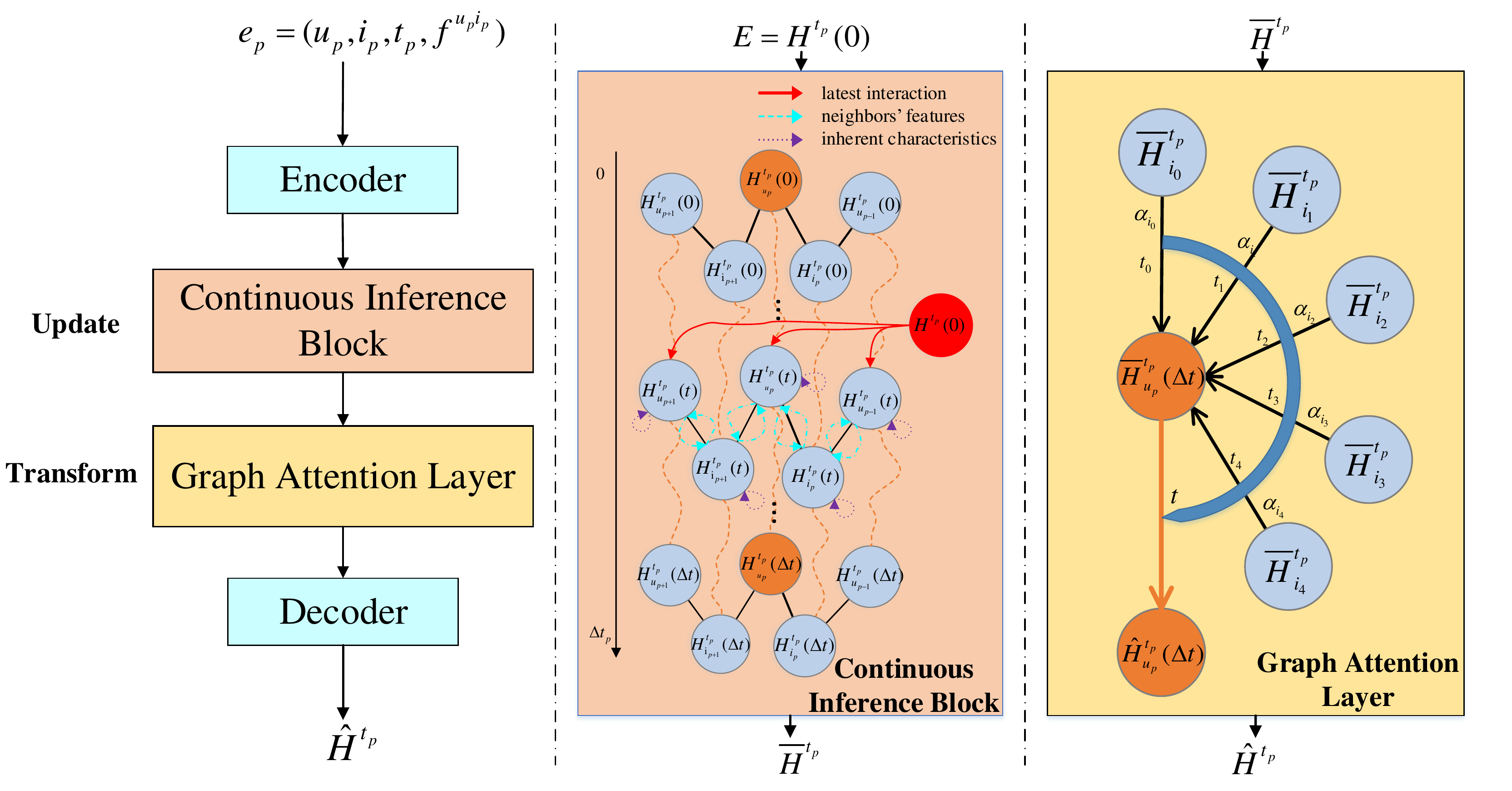}
\caption{ConTIG Framework. The encoder-decoder framework (left) consists of a continuous inference block (middle) and a graph attention layer (right). The former contains a neural ODE defined with three-fold factors (i.e., latest interaction, neighbors' feature and inherent characteristics) to update the node embeddings according to the output of encoder. The latter estimates the future embedding of nodes by aggregating observed historical temporal interaction information.} 
\label{Fig.2}
\end{figure*}

Fig.\ref{Fig.2} provides the framework of our proposed ConTIG. The encoder-decoder framework (Fig.\ref{Fig.2} left) consists of two modules: the update module and the transform module. In the update module (Fig.\ref{Fig.2} middle), a continuous inference block is used to update the embeddings of the interacting nodes in a continuous setting. In the transform module (Fig.\ref{Fig.2} right), a graph attention layer is applied to estimate the future embedding of nodes by capturing observed historical temporal interaction information. When an interaction message $e_p = (u, i, t_p, f^{u i}) $ comes, we first employ a neural encoder to project the latest interaction message $e_q = (u, r_u, \tau_u, f^{u r_u}) $ of input interacting node into a latent space, $i.e., E=f(X)$. Afterwards, treating $E$ as the initial value $H^{t_p} \left ( 0 \right ) $ in continuous inference block, we utilize a neural ordinary differential equations (ODE) to learn the nodes' continuously changing state trajectories at a certain time interval $\left [ 0, \Delta t \right ]$, and the embeddings of nodes $\bar{H} ^{t_{p-1}} _u$ is updated as $\bar{H} ^{t_p} _u$ by the node embeddings obtained in nodes' state trajectory at ending time $H^{t_p} \left ( \Delta t \right )$. Then, selecting $k$ observed historical neighbors of current nodes $u$, we introduce a self-attention mechanism on graphs to convert these interactions information to generate future node embedding $\hat{H}^t_u$. Finally, the decoder uses the future node embeddings $\hat{H}^t_u$ for downstream tasks (i.e., temporal link prediction, temporal node recommendation and dynamic node classification). 

\subsection{Encoder-Decoder} \label{Sec 3.3}
The proposed ConTIG adapts an encoder-decoder architecture (Fig.\ref{Fig.2} left). For each interaction, we first project the features into a latent space to generate the initial hidden representation of nodes. After learning node embeddings at current timestamp, we finally design a decoder for the specific downstream tasks.

Before the encoder, we initialize the latest interactive node $r_v$ and timestamp $\tau_v$ as $0$ for all node $v \in \mathrm{V}$ (line 1 in Algorithm \ref{Alg.1}).

In the encoder, to learn the temporal pattern of each interaction, we use a time encoding function in \cite{ftrl_nips_2019} to obtain a continuous functional mapping $F_T: T \rightarrow \mathbb{R}^{d_T}$ from time domain to the $d^T$-dimensional vector space, which projects the timestamps of interactions into the continuous-time space. For any given time $t$, we will generate its time embeddings as follows:
\begin{equation}\label{Eq.1}
\begin{split}
F_T(t) = \frac{1}{\sqrt{d_T} }  [ cos\left ( \omega _1 t \right ) &, sin \left ( \omega _1 t \right ),\\ 
... &, cos\left ( \omega _{d_T} t \right ), sin\left ( \omega _{d_T} t \right )    ],
\end{split}
\end{equation}
where $\omega_1, \omega_2, ... , \omega_{d^T}$ are trainable parameters. To learn informative node representations $H^{t_p}$ at timestamp $t_p$, we concatenate the latest node embedding of the interactive node $\bar{H}^{t_{p-1}}_u$ and it latest interactive node $\bar{H}^{t_{p-1}}_{r_u}$, the last interaction message feature $f^{ur_u}$ and the time embeddings of the time interval $F_T(\Delta t)$ to represent new node features of $u$ at timestamp $t$ (line 3 in Algorithm \ref{Alg.1}). Here, as we focus on the interval between current interaction and its latest interaction, we use the time embedding $F_T(\Delta t)$. Afterwards, we adopt a fully connected layer as an encoder (line 4 in Algorithm \ref{Alg.1}), and the hidden representations can be defined as follows:
\begin{align}\label{Eq.2}
E & = f(\bar{H}^{t_{p-1}}_u \left |  \right |  \bar{H}^{t_{p-1}}_{r_u} \left |  \right |  f^{u r_u} \left |  \right |   F_T(\Delta t)),   
\end{align}
where $\left | \right | $ is the concatenation operation, and $f(\cdot)$ is a linear projection as follows:
\begin{align}
f(x) = xW+b,   
\end{align}
where $W$ and $b$ are learnable parameters.

For the decoder, two fully connected layers are designed for temporal link prediction and temporal node recommendation tasks, and three fully connected layers are designed for dynamic node classification task.

\subsection{Update: Continuous Inference Block} \label{sec 4.3}

In update module (Fig.\ref{Fig.2} middle), to capture the continuous dynamic of node representation, inspired by \cite{cgnn_icml_2020}, we define a ODE with the latest interaction information, neighbor features and inherent characteristics of nodes, and use a neural solver to generate the node state trajectories at a certain time interval $\left [ 0, \Delta t \right ]$. In this way, our method can estimate the possible direction of embedding trajectory for an inactive node (line 5 and 11-16 in Algorithm \ref{Alg.1}. 

We select the interactions between each node in temporal interaction graph $u$ and its last interactive node $r_u$, and divide them into a set $\mathcal{E}_p$ to generate an adjacency matrix $S_p$ describing their relationships. $S_p \in \mathbb{R} ^{\left | \mathcal{V} \right | \times \left | \mathcal{V} \right |}$ is defined by $\mathcal{E}_p$ as follows:
\begin{align}\label{Eq.4}
S_p^{ui} & = \begin{cases}
 1 & \text{ if } (u, r_u)\in \mathcal{E}_p  \\
 0 & \text{ otherwise},
\end{cases}
\end{align}

As the degree of nodes can be very different, we typically normalize the adjacency matrix as $D_p^{-\frac{1}{2}} S_p D_p^{-\frac{1}{2}}$, where $D_p=\mathrm {diag} \left ( \sum_{j}S_p^{ij}  \right ) \in \mathbb{R}^{M \times M }$ is the degree matrix of $S_p$. As such a normalized adjacency matrix always has an eigenvalue decomposition, and the eigenvalues are in the interval $\left [-1, 1\right ]$. To get the positive eigenvalues and make graph learning algorithms stable in practice,  we follow \cite{gcn_iclr_2017} and leverage the following regularized matrix for characterizing graph structures:
\begin{align}\label{Eq.5}
A_p & = \frac{\beta }{2} \left (I_N+D_p^{-\frac{1}{2}} S_p D_p^{-\frac{1}{2} }\right ),
\end{align}
where $\beta \in \left (0, 1 \right)$ is a hyperparameter, $I_N \in \mathbb{R}^{N \times N }$ is the identity matrix, $N=\left | \mathrm{V} \right |$, and eigenvalues of $A_p$ are in $\left [0, \beta \right]$.

Afterwards, to capture the complex non-linear dynamic of node embeddings, we assume that there are three possible factors affecting the node states: 1) latest interaction information exhibiting the latest nodes states; 2) neighbors’ features affecting the change of node states; 3) the inherent characteristics of nodes determining the influence of the aforementioned two factors. Based on these considerations, we define an ODE in $\mathcal{E}_p$ to solve the node embedding trajectories, which adaptively fuse them with a gated mechanism. Here, we treat the encoder $E_{t_p}$ as the initial value of ODE $H^{t_p}(0)$, i.e., $H^{t_p}(0) = E_{t_p}$. Then we use a differential equation defined as follows to learn the continuous node representations in the interval of interactions:
\begin{align}\label{Eq.6}
\frac{\mathrm{d} H^{t_p}(t)}{\mathrm{d} t} =  \underset{\text{latest interaction}}{\underbrace{z_l \odot H^{t_p}(0)}}  + \underset{\text{neighbor features}}{\underbrace{z_n \odot A_pH^{t_p}(t)}} - \underset{\text{inherent characteristics}}{\underbrace{z_i \odot H^{t_p}(t)}},
\end{align}
where $H^{t_p}(0)$ represents the latest interaction information of nodes, $A_pH^{t_p}(t)$ represents the influence from neighbors, $-H^{t_p}(t)$ represents the inherent characteristics of nodes, $\odot$ denotes the element-wise product operation. $z_l$, $z_n$, and $z_i$ represents three gates, respectively. They are computed as:
\begin{align}\label{Eq.7}
z_l & = \sigma (W_l H^{t_p}(0) + b_l),\\
z_n & = \sigma (W_n H^{t_p}(0) + b_n),\\
z_i & = \sigma (W_i H^{t_p}(0) + b_i),
\end{align}
where $W_l, W_n, W_i$ and $b_l, b_n, b_i$ are trainable parameters, and $\sigma (\cdot)$ is the sigmoid activation function to normalize the output into $[0, 1]$. The gated fusion mechanism can adaptively fuse three factors according to their importance calculated by the initial value of ODE. In addition, the updating process starts at the timestamp of latest interaction of the node. To this end, following \cite{tgn_icmlgrl_2020}, when an interactive message comes, we use the latest interaction information to update node embeddings and save the current interaction message for the next time the node appears. Meanwhile, due to the integration of time encoding in $H^{t_p}(0)$, the temporal behaviors of nodes could be captured. As a result, we can leverage the aforementioned ODE to learn the nodes' state trajectories at the interval $\left [ 0, \Delta t \right ]$, and the three main factors mentioned above jointly captures the continuous dynamic of node representations. Then the node representations are updated by ODE solver as follows:
\begin{align}\label{Eq.10}
H^{t_p} (\Delta t) = \mathrm {ODESolve}(g(t), H^{t_p}(0), \Delta t)
\end{align}
where $g(t) = \frac{\mathrm{d} H^{t_p}(t)}{\mathrm{d} t}$.

Finally, we use the hidden state at the end time $H (\Delta t_p)$ to update the previous embedding memory $\bar{H}^{t_{p-1}}$ as $\bar{H}^{t_{p}}$ (i.e., $\bar{H}^{t_{p}} = H (\Delta t_p)$) (line 5 in Algorithm \ref{Alg.1}). 

\subsection{Transform: Graph Attention Layer} \label{sec 4.4}
After capturing the continuous dynamic of node representations from latest interactions sets and updating the embeddings of nodes, a graph attention layer in transform module (Fig.\ref{Fig.2} right) is applied to convert the historical observed interaction features of nodes to generate future representations (Line 6 and 17-24 in Algorithm \ref{Alg.1}).

In this module, for the current interaction $e$ connected by $u$ and $v$ at time $t$, their temporal neighbors and the interaction information between them are token as input. We introduce a self-attention mechanism to distinguish different neighbors, and take account of the structural information with temporal information \cite{tgat_iclr_2020}. For node $u$ at time $t$, we consider its neighbors $N(u, t) = \{i_0,...,i_{k-1}\}$, where the interactions between $u$ and $i_j\in N(u, t)$ occurred at time $t_j$ prior to time $t$, and the sampling process of temporal neighbors of node $i$ is the same as $u$. Then we take the node information with time $t$ encoding $Z^t_u = \bar{h}^t_u \left |  \right |  F_T(t)$ and the neighborhood information with the time delta $t-t_j$ encoding $Z^t_N = \bar{H}^t_{N(u, t)} \left |  \right |   f^{uN(u, t)} \left |  \right |  F_T(t-t_j)$ as the input of attention, where the time delta $t-t_j$ is between current interaction $e$ and the interaction of $u$ and its neighbors $i_j \in N(u, t)$. In attention, three different linear projections is used to obtain the query $Q_t$, key $K_t$ and value $V_t$:
\begin{align}
q_t = Z^t_u W_Q,K_t = Z^t_N W_K,V_t = Z^t_N W_V,
\end{align}
where $W_Q, W_K, W_V $ are the weight matrices. The attention weights $\alpha_j$ is given by:
\begin{align}
\alpha_j = \frac{\exp (q^{\mathrm {T}}K_j/\sqrt{d})}{\textstyle \sum_{c}\exp \left ( q^{\mathrm {T}} K_c /\sqrt{d}  \right )}, 
\end{align}
where the attention weight $\alpha_j$ reveals how node $u$ attends to the features of node $i_j$ within the topological structure defined as $N(u, t)$ after accounting for their interaction time with $i_j$. Therefore, the layer produces the time-aware representation of node $u$ at time $t$, $\hat{h}_u^t$, which represents the future hidden representations generated by historical observed interactions as follows:
\begin{align}
\hat{h}_u^t & = \sum_{j=0}^{k-1} \alpha_j V_t,
\end{align}

To stabilize the learning process and improves performances, we extend the self-attention mechanism to be multi-head ones \cite{attention_nips_2017}. Specifically, we concatenate $M$ parallel attention mechanisms with different learnable projections:
\begin{align}
{\hat{h}_u^t} & = \left |  \right | _{m=1}^{M} \left \{ \sum_{j=0}^{k-1} \alpha^{(m)}_j V^{(m)}_t \right \}, 
\end{align}
where $V^{(m)}_t$ represent the value with different projections in the $m^{th}$ head attention, and $\alpha^{(m)}_j$ represent the attention weight calculated by the query and key with different projections in the $m^{th}$ head attention.

Finally, the future node embeddings $\hat{H}_u^{t_p}$ is generated by calculating $\hat{h}_u^{t_p}$ for each node in interactions (line 6 in Algorithm \ref{Alg.1}). The latest interactive node $r_u$ and $r_i$ are updated as $i$ and $u$, respectively. In addition, the latest interactive timestamps $\tau_u$ and $\tau_i$ are both updated as $t_p$ (Line 7-8 Algorithm \ref{Alg.1}).

\subsection{Binary Cross-Entropy Loss Function}
In this work, we adopt time-sensitive link prediction binary cross-entropy loss function to learn ConTIG’s parameters. The binary cross-entropy loss function is defined as follows and our goal is to maximize this likelihood function:
\begin{equation}
\begin{split}
\mathcal{L} = \sum_{(u_p, i_p, t_p)\in \mathcal{E } } &- \log \sigma ({-\hat{h}_{u_p}^{t_p}}^{\mathrm {T}} \hat{h}_{i_p}^{t_p}) \\
& -Q\mathbb{E}_{i_q\sim P(i)}\log{\sigma ({-\hat{h}_{u_p}^{t_p}}^\mathrm {T} \hat{h}_{i_q}^{t_p}}),   
\end{split}
\end{equation}
where the summation is over the observed edges on $u_p$ and $i_p$ that interact at time $t_p$, $Q$ is the number of negative samples and $P(i)$ is the negative sampling distribution over the node space. 

\begin{algorithm}[!t]
\caption{Continuous Representation Learning on Temporal Interaction Graphs}\label{Alg.1}
{\bf Input:} 
Edge Stream $e=(u, i, t, f^{u i}) \in \mathcal{E}$ in temporal interaction graph $\mathcal{G}=(\mathcal{V}, \mathcal{E})$\\
{\bf Output:} 
Embedding results ${\hat{h}_u^t}$ for all nodes
\begin{algorithmic}[1]
\State $\bar{h}_v^0 \gets x_v, r_v \gets 0, \tau_v \gets 0, \forall v \in  \mathcal{V}$
\For{$e_{p}=(u, i, t_p, f^{u i})$ \textbf{in} $\mathcal{E}$}
    \State Learning $ F_T(t_p-\tau_v) $ by Eqn.\ref{Eq.1}
    \State $E_v^{t_p}=f(\bar{h}_v^{t_{p-1}} \left |  \right |  \bar{h}_{r_v}^{t_{p-1}} \left |  \right |  f^{vr_v} \left |  \right |   F_T(t_{p}-\tau_v)), \forall v \in \mathcal{V}$
    \State $\bar{h}^{t_p}_u, \bar{h}^{t_p}_i \gets$ Continuous Inference Block $(E^{t_p}, u, i)$
    \State $\hat{h}^{t_p}_u, \hat{h}^{t_p}_i \gets$ Graph Attention Layer $(\bar{h}^{t_p}_u, \bar{h}^{t_p}_i, u, i, t_p)$
    \State $r_u \gets i, \tau_u \gets t_p$
    \State $r_i \gets u, \tau_i \gets t_p$
\EndFor
\State return $\bar{h}_v^{t}, \forall v \in  \mathcal{V}$
\Procedure{Continuous Inference Block}{$E^{t_p}, u, i$}
    \State $h_v(0) \gets E^{t_p}_v, v \in \mathcal{V}$
    \State $(v, r_v), \forall v \in \mathcal{V}$ generates $A_p$ by Eqn.\ref{Eq.4}-\ref{Eq.5}  
    \State $h_v(\Delta t) \gets \mathrm{ODESolve}(g(t), h_v(0), \Delta t)$, $\forall v \in  \mathcal{V}$
    \State \Return $h_u(\Delta t), h_i(\Delta t)$
\EndProcedure
\Procedure{Graph Attention Layer}{$\bar{h}^t_u, \bar{h}^t_i, u, i, t$}
    \For{$v$ \textbf{in} $\left \{u, i \right \}$}
       \State Sample $k$ temporal neighbors $N(v, t)$
       \State $z^t_v \gets \bar{h}^t_v \left |  \right |  F_T(t), $ 
       \State $Z^t_{N(v, t)} \gets \bar{H}^t_{N(v, t)} \left |  \right |   f^{vN(v, t)} \left |  \right |  F_T(t-t_j)$
       \State $h^t_{N(v, t)} \gets \mathrm{Attn}(z^t_v, Z^t_{N(v, t)}, Z^t_{N(v, t)})$
       \State $\hat{h}^t_v \gets f(h^t_{N(v, t)} \left | \right | \bar{h}^t_v)$
    \EndFor
    \State \Return $\hat{h}^t_u, \hat{h}^t_i$
\EndProcedure
\end{algorithmic}
\end{algorithm}

\subsection{Complexity Analysis}
The time complexity of our proposed method mainly consists of two portions.

First, for the update module (i.e., continuous inference block), we consider that both adjacency matrices are stored as sparse matrices. And the runtime of the ODE solver depends on the length on the time interval (i.e., the end time of ODE solver) $\Delta t$ and the complexity of the dynamics. Then, the time complexity of the continuous inference block is $O(\Delta t \left | \mathcal{E} \right | )$. 

Second, for the transform module (i.e., graph attention layer), since the
masked self-attention operation is parallelizable, as suggested by \cite{attention_nips_2017}. The per-batch time complexity of the graph attention layer with $m$ heads can be expressed as $O(mk)$, where $k$ is the average neighborhood size. Since the batch is divided by edges, the time complexity of the graph attention layer is $O( m k \left | \mathcal{E}  \right | )$, where $m \ll \left | \mathcal{E} \right |$ and $k \ll \left | \mathcal{E}  \right |$.

Therefore, the time complexity of ConTIG is $O((\Delta t + m k) \left | \mathcal{E}  \right |) \approx O(C \left | \mathcal{E}  \right |)$, where $C$ is a constant and is relative to the runtime of the ODE Solver.

\section{Experiments}
In this section, we will utilize four networks to compare the performance of our model with four static methods and five dynamic methods. We conduct three main tasks, link prediction, node recommendation and node classification, to evaluate the influence of introducing temporal information and learn continuous dynamic of node embedding trajectories.

\subsection{Experimental Setup}

\textbf{Datasets} We evaluate the performance of ConTIG on temporal link prediction, node recommendation and dynamic node classification tasks with four public datasets, where three datasets are user-item networks selected by \cite{jodie_kdd_2019} and one dataset is e-mail network. The statistics of the four datasets are summarized in Table \ref{tab.1}.

\begin{itemize}
    \item \textbf{Wikipedia}\footnote{http://snap.stanford.edu/jodie/wikipedia.csv}. The dataset describes the interactions between active users and pages they edit the contents with unique timestamps and the dynamic labels indicating whether a user is banned from editing. 

    \item \textbf{Reddit}\footnote{http://snap.stanford.edu/jodie/reddit.csv}. The dataset describes the interactions between active users and the posts they submit on subreddits and the dynamic labels indicating if users are banned from posting. 
    
    \item \textbf{Mooc}\footnote{http://snap.stanford.edu/jodie/mooc.csv}. The dataset describes the interactions between students and MOOC online course, e.g., viewing a video, submitting an answer, etc.
    
    \item \textbf{CollegeMsg}\footnote{https://snap.stanford.edu/data/CollegeMsg.html}. The dataset is an online social network at the University of California, and describes the interaction between users by sending private message at different timestamps. The dataset is without node label and edge features. 
\end{itemize}

\textbf{Baselines} To evaluate the performance of ConTIG, we compare our method with state-of-the-art graph embedding methods on both static and dynamic graphs. 

\textbf{Static graph embedding methods}: GAE and VGAE \cite{gae_arxiv_2016} learn latent representations by graph auto-encoder. GraphSAGE \cite{graphsage_nips_2017} and GAT \cite{gat_iclr_2018} are GNNs that capture the dependence of graphs via message passing between the nodes of graphs. 

\textbf{Dynamic graph embedding methods}: CTDNE \cite{ctdne_www_2018} defines the temporal random walk requiring the walks to obey the temporal order. JODIE \cite{jodie_kdd_2019} learns to generate embedding trajectories of all users and items from temporal interactions by update and project operations. DyRep \cite{dyrep_2019_iclr} use deep recurrent architecture and attention mechanism to effectively model fine-grained temporal dynamic. TGAT \cite{tgat_iclr_2020} introduces a self-attention mechanism and a functional time encoding technique to learn the time-feature interactions. TGN \cite{tgn_icmlgrl_2020} is a generic inductive framework operating on continuous-time dynamic graphs \cite{survey_2020_jmlr} represented as a sequence of events.

\begin{table}[!t]
\renewcommand{\arraystretch}{1.3}
\caption{Statistics of the datasets.}
\label{tab.1}
\centering
\begin{tabular}{ccccc}
\hline
\bfseries Datasets & \bfseries $\left |\mathcal{V} \right |$ & \bfseries $\left |\mathcal{E}  \right |$ & \bfseries Feature & \bfseries Label\\
\hline\hline
Wikipedia & 9,227 & 157,474  & 172 & 2\\
Reddit & 10,984 & 672,447  & 172 & 2\\
Mooc & 7,144 & 411,749 & 0 & 2 \\
CollegeMsg & 1,899 & 59,835 & 0 & 0 \\
\hline
\end{tabular}
\end{table}

\begin{table*}[!t]
\renewcommand{\arraystretch}{1.3}
\caption{ROC AUC(\%) and Average Precision(\%) for the transductive temporal link prediction on Wikipedia, Reddit, Mooc and CollegeMsg. The means and standard deviations are computed ten runs.}
\label{tab.2}
\centering
\begin{tabular}{llcccccccc}
    \toprule
    \multirow{2}*{Task} & \multirow{2}*{Methods} & \multicolumn{2}{c}{Wikipedia} & \multicolumn{2}{c}{Reddit} &
    \multicolumn{2}{c}{Mooc} &
    \multicolumn{2}{c}{CollegeMsg} \\
    \cline{3-10}
    ~ & ~ &  AUC & AP & AUC & AP & AUC & AP & AUC & AP \\
    \midrule
    \multirow{8}*{\rotatebox{90}{Transductive}} & GAE & $91.47\pm0.3$ & $91.12\pm0.1$ & $95.87\pm1.2$ & $96.57\pm1.0$ & $87.89\pm0.6$ & $90.70\pm0.3$ & $73.15\pm1.5$ & $70.00\pm1.17$ \\
    ~ & VGAE        & $82.43\pm1.6$ & $82.50\pm4.0$ & $92.70\pm0.4$ & $91.53\pm0.7$ & $88.21\pm0.6$ & $\textbf{91.00}\pm0.3$ & $74.07\pm0.9$ & $70.66\pm1.0$  \\
    ~ & GraphSAGE   & $92.00\pm0.3$ & $92.34\pm0.3$ & $97.75\pm0.1$ & $97.85\pm0.1$ & $56.17\pm0.3$ & $60.63\pm0.2$ & $62.38\pm1.3$ & $62.48\pm0.9$ \\
    ~ & GAT       & $92.76\pm0.5$ & $93.17\pm0.5$ & $97.90\pm0.1$ & $97.07\pm0.1$ & $67.24\pm0.1$ & $66.66\pm0.8$ & $78.09\pm0.5$ & $75.97\pm0.7$ \\
    ~ & CTDNE       & $ 82.36\pm0.7 $ & $ 80.86\pm0.7 $ & $ 85.32\pm2.0 $ & $ 87.31\pm1.4 $ & $ 88.97\pm2.6 $ & $ 89.27\pm2.0 $ & $ 81.88\pm0.7 $ & $ 80.25\pm0.8 $ \\
    ~ & JODIE       & $ 94.94\pm0.3 $ & $ 94.65\pm0.6 $ & $ 97.62\pm0.2 $ & $ 97.07\pm0.4 $ & $ 79.75\pm2.8 $ & $ 74.85\pm3.1 $ & $59.85\pm6.0$ & $54.50\pm4.4$ \\
    ~ & DyRep       & $ 94.22\pm0.2 $ & $ 94.63\pm0.2 $ & $ 98.01\pm0.1 $ & $ 98.05\pm0.1 $ & $ 80.57\pm2.1 $ & $ 77.30\pm2.2 $ & $54.75\pm6.8$ & $51.89\pm4.8$ \\
    ~ & TGAT        & $ 94.99\pm0.3 $ & $ 95.29\pm0.2 $ & $ 98.07\pm0.1 $ & $ 98.17\pm0.1 $ & $ 66.02\pm1.0 $ & $ 63.82\pm0.9 $ & $81.05\pm0.6$ & $79.16\pm0.6$ \\
    ~ & TGN         & $ 98.42\pm0.1 $ & $ 98.50\pm0.1 $ & $ 98.69\pm0.1 $ & $ 98.73\pm0.1 $ & $ 89.07\pm 1.6 $ & $ 86.96\pm2.1 $ & $85.06\pm5.9$ & $85.38\pm6.4$ \\
    ~ & ConTIG      & $ \textbf{98.50}\pm0.2$ & $ \textbf{98.62}\pm0.2$ &  $\textbf{98.71}\pm0.3 $ & $ \textbf{98.75}\pm0.3 $ &  $\textbf{90.34}\pm1.6 $ & $ 88.87\pm1.9 $ &  $\textbf{90.11}\pm1.2$ & $\textbf{90.54}\pm1.2$ \\
    \cline{1-10}
    \multirow{8}*{\rotatebox{90}{Inductive}}    & GraphSAGE   & $88.60\pm0.3$ & $88.94\pm0.5$ & $94.28\pm0.4$ & $94.51\pm0.1$ & $53.68\pm0.4$ & $55.35\pm0.4$ & $49.64\pm1.5$ & $51.83\pm0.8$ \\
    ~ & GAT       & $89.11\pm0.5$ & $89.82\pm0.4$ & $94.30\pm0.4$ & $94.58\pm0.3$ & $53.43\pm2.1$ & $54.80\pm0.9$ & $68.98\pm1.2$ & $66.22\pm1.2$ \\
    ~ & JODIE       &
    $ 92.75\pm0.3 $ & $ 93.11\pm0.4 $ & $ 95.42\pm0.2 $ & $ 94.50\pm0.6 $ & $ 81.43\pm0.8 $ & $ 76.82\pm1.4 $ & $ 51.59\pm3.2 $ & $ 50.02\pm2.2 $ \\
    ~ & DyRep       & $ 91.03\pm0.3 $ & $ 91.96\pm0.2 $ & $ 95.79\pm0.5 $ & $ 95.75\pm0.5 $ & $ 82.06\pm1.7 $ & $ 79.17\pm1.6 $ & $ 49.05\pm4.1 $ & $ 49.30\pm2.6 $ \\
    ~ & TGAT        & $ 93.37\pm0.3 $ & $ 93.86\pm0.3 $ & $ 96.46\pm0.1 $ & $ 96.61\pm0.2 $ & $ 69.09\pm0.8 $ & $ 67.65\pm0.7 $ & $ 72.27\pm0.5 $ & $ 72.53\pm0.6 $ \\
    ~ & TGN         & $ 97.72\pm0.1 $ & $ 97.83\pm0.1 $ & $ 97.54\pm0.1 $ & $ 97.63\pm0.1 $ & $ \textbf{89.03}\pm1.6 $ & $ \textbf{86.70}\pm2.0 $ & $ 78.54\pm3.9 $ & $ 80.77\pm3.7 $ \\
    ~ & ConTIG      & $ \textbf{98.44}\pm0.2 $ & $ \textbf{98.41}\pm0.2 $ & $ \textbf{98.26}\pm0.3 $ & $ \textbf{98.31}\pm0.2 $ & $ 86.77\pm2.0 $ & $ 85.44\pm2.0 $ & $ \textbf{83.63}\pm0.9 $ & $ \textbf{85.37}\pm0.9 $ \\
    \bottomrule
\end{tabular}
\end{table*}

\textbf{Parameter Settings} For parameter settings, the dimension of both node representations and time representations are set as 172. The optimizer is Adam algorithm, learning rate is 0.0001, and dropout probability is 0.1. In continuous inference block, parameter $\beta$ in adjacency matrix regularization is 0.95, and the end time of ODE is set as 1.0 instead of the real interval $\Delta t$. In graph attention layer, the number of selected neighbors $k$ is set as 15 and heads $M$ in attention is set as 2, and the negative sampling distribution $P(i)$ is uniform distribution. 

\textbf{Settings for Baselines} For static baselines, we adopt the same baseline training procedures as in \cite{tgat_iclr_2020}. We refer to the PyTorch geometric library for implementing the GAE and VGAE baselines, and develop off-the-shelf implementation for GraphSAGE and GAT by referencing their original implementations to accommodate the temporal setting and incorporate edges features. For dynamic baselines, we use the open-source implementations for CTDNE, and use the source code released by the authors to implementing TGAT and TGN, where we implement JODIE and DyRep in the version of TGN in PyTorch.

\textbf{Implementation} All our experiments are implemented on a 32g-MEM Ubuntu 20.04 server with Intel(R) Core(TM) i7-9700K CPU @ 3.60GHz and NVidia(R) 2080Ti GPUs. All code is implemented in Pytorch version 1.5.1. 

\subsection{Temporal Link Prediction}
The goal of the temporal link prediction task is to predict the probability that there exists a link between the two nodes given two nodes and a certain timestamp in the future. For this task, we evaluate our method on both the transductive and inductive setting. In the transductive task, we predict the links between nodes observed during training. In the inductive task, we predict the links between new nodes which have not been observed in the past. Our model is tuned on the validation set and we report the average precision (AP) on the test set. We divide the training, validation, and testing sets into a 70\%-15\%-15\% split. After the division, the nodes that do not appear in the training set are considered as new nodes. The model is trained by temporal link prediction.

The results comparison between our method and baseline methods in temporal link prediction tasks are shown in Table \ref{tab.2}. We observe that: 1) static graph embedding methods including GAE, VGAE, GraphSAGE and GAT, perform worse than those baselines modeling temporal interaction information. Because most of the interactions in the real-world networks are time-stamped; 2) among the dynamic graph embedding works, both temporal neighbor information (i.e., CTDNE, TGAT) and latest interaction information (i.e., JODIE, DyRep) methods perform worse than fusing methods (i.e., TGN, ConTIG); and 3) ConTIG achieves the best prediction performance on the datasets Wikipedia, Reddit, and CollegeMsg for both transductive and inductive settings, and the second best performance on Mooc for the inductive task. This observation demonstrates the advantage of our proposed method compared to existing methods. In fact, by considering the inherent node properties and modeling the three important factors (i.e., latest interaction, neighbor features and inherent characteristics) in temporal interaction graphs, our method can capture the complex non-linear dynamics of node representations effectively, thereby achieving superb performance.

\subsection{Temporal Node Recommendation}

\begin{figure*}[!t]
\centering
\subfloat[Recall@K on Wikipedia]{\includegraphics[width=2.2in]{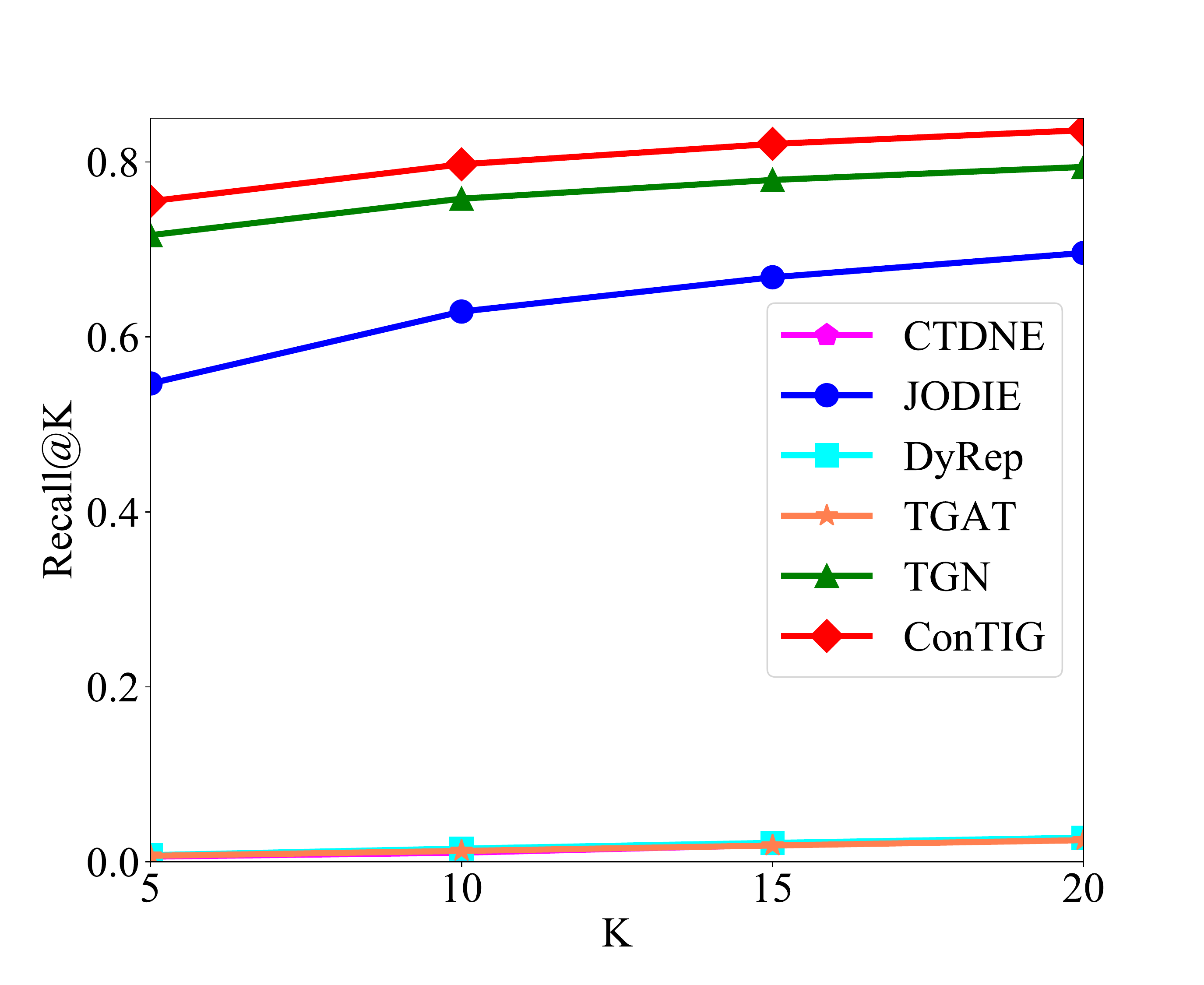}}
\hfil
\subfloat[Recall@K on Reddit]{\includegraphics[width=2.2in]{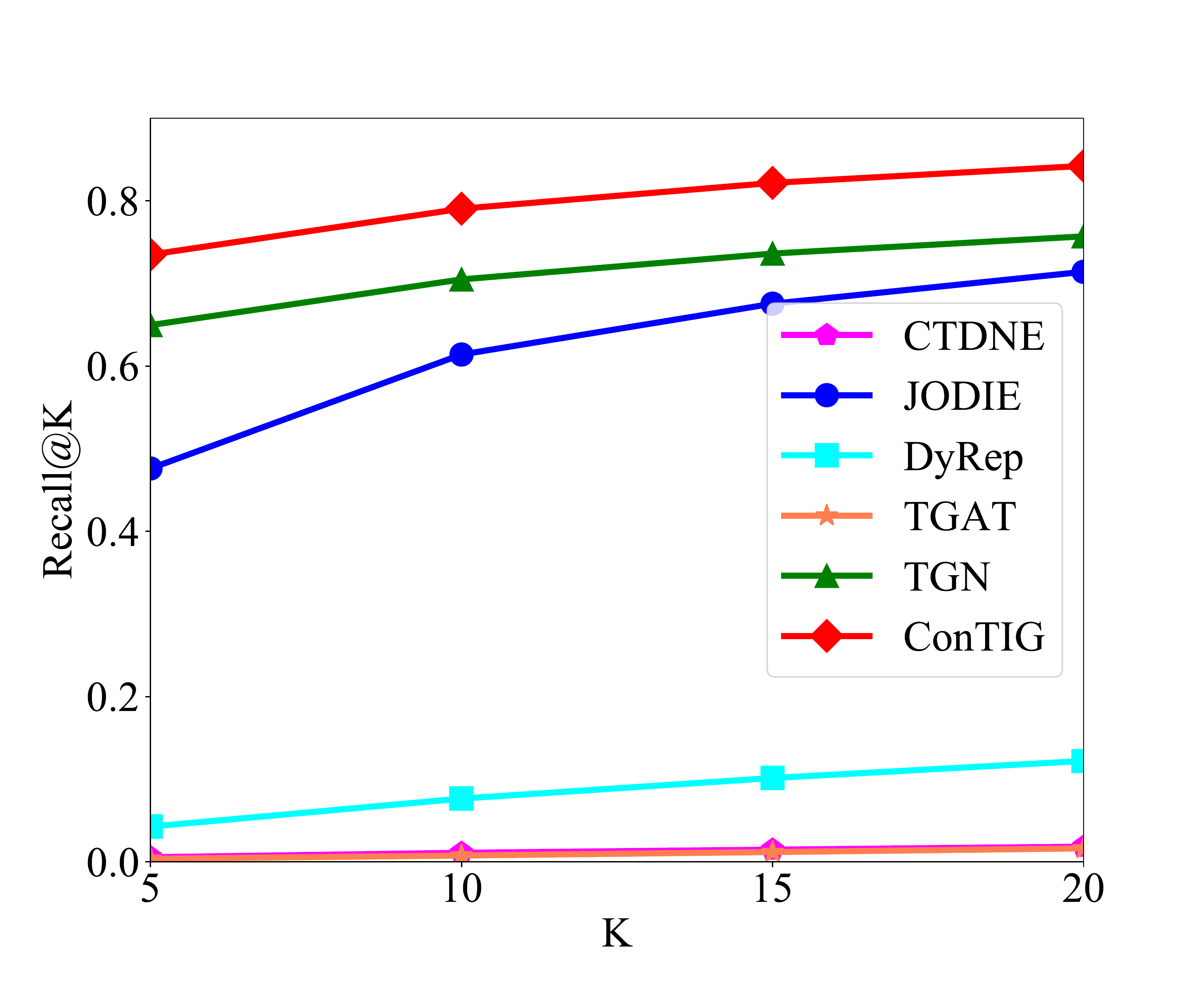}}
\hfil
\subfloat[Recall@K on Mooc]{\includegraphics[width=2.2in]{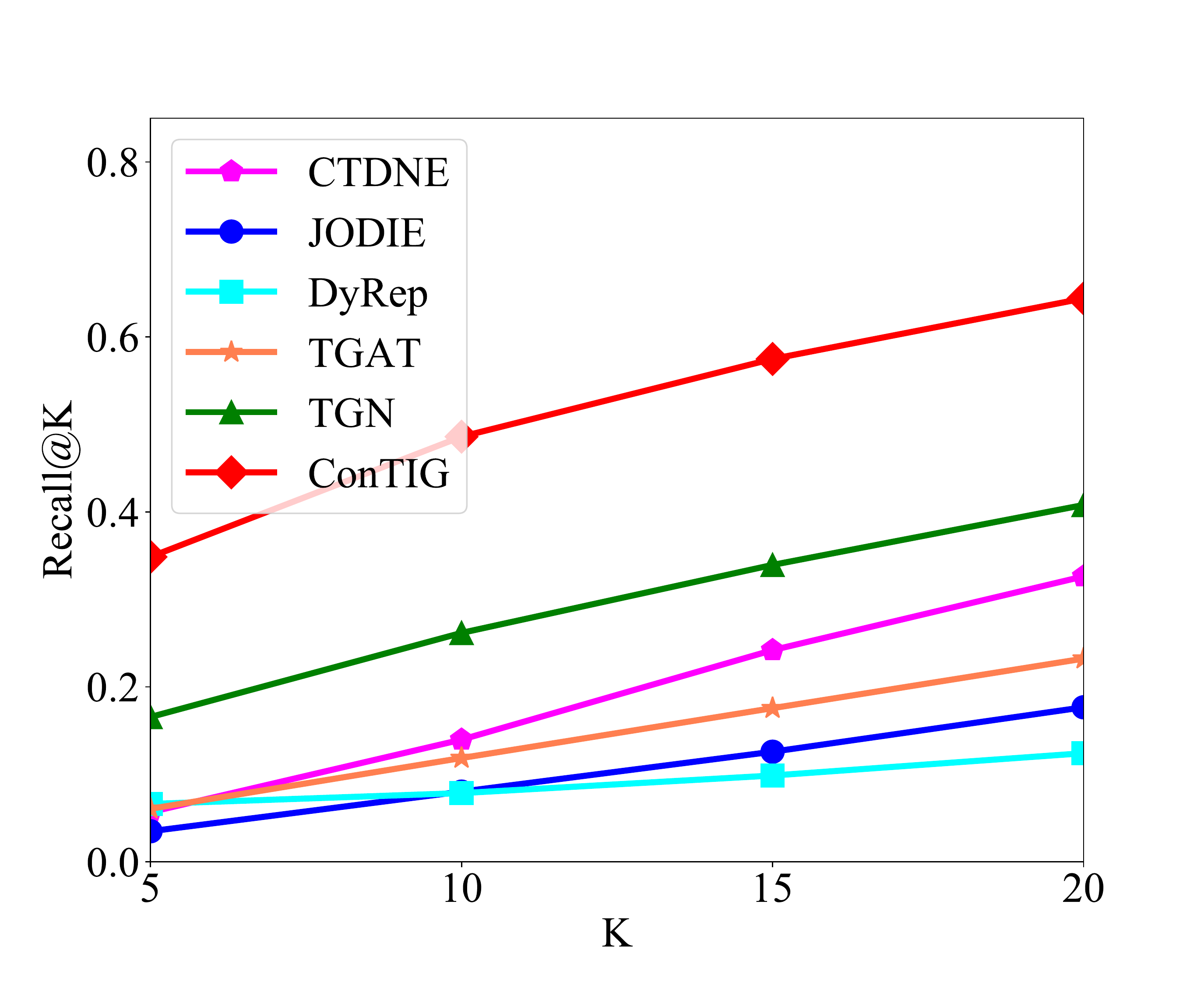}}
\caption{Recall@K for the transductive temporal node recommendation on Wikipedia, Reddit and Mooc.}
\label{Fig.3}
\end{figure*}

The goal of temporal node recommendation task is to predict the top-K possible neighbors of node $u$ at $t$ given the node $u$ and future timestamp $t$. This task is also used to evaluate the performance of temporal network embedding methods. For this task, we evaluate all methods on transductive setting, each method outputs the user $u$’s preference scores over all the items at time in test set. We sort scores in a descending order and record the rank of the paired node $u$ and $i$. We evaluate the task on three user-item networks (i.e., Wikipedia, Reddit and Mooc), where CollegeMsg dataset is not included because it is not a user-item network. Our evaluation metric in this task is Recall@K, where $K \in \left \{5, 10, 15, 20 \right \}$.

The results comparison between our method and baseline methods in temporal node recommendation task is shown in Fig.\ref{Fig.3}, showing that our model ConTIG perform better than all the baselines. Compared with the best competitors (i.e., TGN), the recommendation performance of ConTIG improves by 110.51\% in terms of Recall@5 on Mooc. On Wikipedia and Reddit, the improvement is 3.25\% and 13.13\% with respect to Recall@5. These significant improvements verify that the differential equation fused with three factors (i.e., latest interaction, neighbor features and inherent characteristics) proposed in ConTIG is capable of learning the trend of node state trajectories in the network. Additionally, the significant improvement of ConTIG benefits from the combination of continuous node state modeling in update module and dynamic sub-graph structure capturing in transform module on node embeddings, which is good for the down-stream node recommendation task.

\subsection{Dynamic Node Classification}

\begin{table}
\renewcommand{\arraystretch}{1.3}
\caption{ROC AUC(\%) for the transductive dynamic node classification on Wikipedia, Reddit and Mooc. The means and standard deviations are computed ten runs. } 
\label{tab.3}
\centering
    \begin{tabular}{lccc}
    \toprule
      & Wikipedia & Reddit & Mooc \\
    \midrule
    CTDNE        & $84.86\pm1.5$  & $54.38\pm7.5$ & $71.84\pm1.0$ \\
    JODIE        & $84.40\pm0.9$ & $61.51\pm1.2$ & $70.03\pm0.5$ \\
    DyRep        & $83.25\pm0.5$ & $60.86\pm1.7$ & $64.64\pm1.4$ \\
    TGAT         & $84.41\pm1.5$ & $65.98\pm1.6$ & $65.79\pm0.5$ \\
    TGN          & $\textbf{87.56}\pm0.7$ & $69.78\pm0.8$ & $63.93\pm0.3$ \\
    ConTIG       & $87.13\pm0.6$ & $\textbf{69.99}\pm0.5$ & $\textbf{73.56}\pm0.4$\\
    \bottomrule
    \end{tabular}
\end{table}

The goal of dynamic node classification task is to predict the state label of user given the user, item and future timestamp. For this task, we evaluate our method on transductive setting, predicting the state labels of users who have been observed during training. We evaluate the task on three datasets with dynamic node labels (i.e., Wikipedia, Reddit and Mooc), where CollegeMsg dataset is not included because there is no node label. Specifically, we train a decoder after the the model trained by the temporal link prediction task. Our evaluation metric in this task is the area under the ROC curve (AUC).

The results comparison between our method and baseline methods in dynamic node classification tasks are shown in Table \ref{tab.3}. Again, our algorithm achieves the best or comparable performance comparing with existing dynamic graph embedding approaches, demonstrating its effectiveness for the down-stream node classification task.


\subsection{Performance on Long-interval Interactions}

\begin{figure}[!t]
\centering
\subfloat[Wikipedia ]{\includegraphics[width=3.5in]{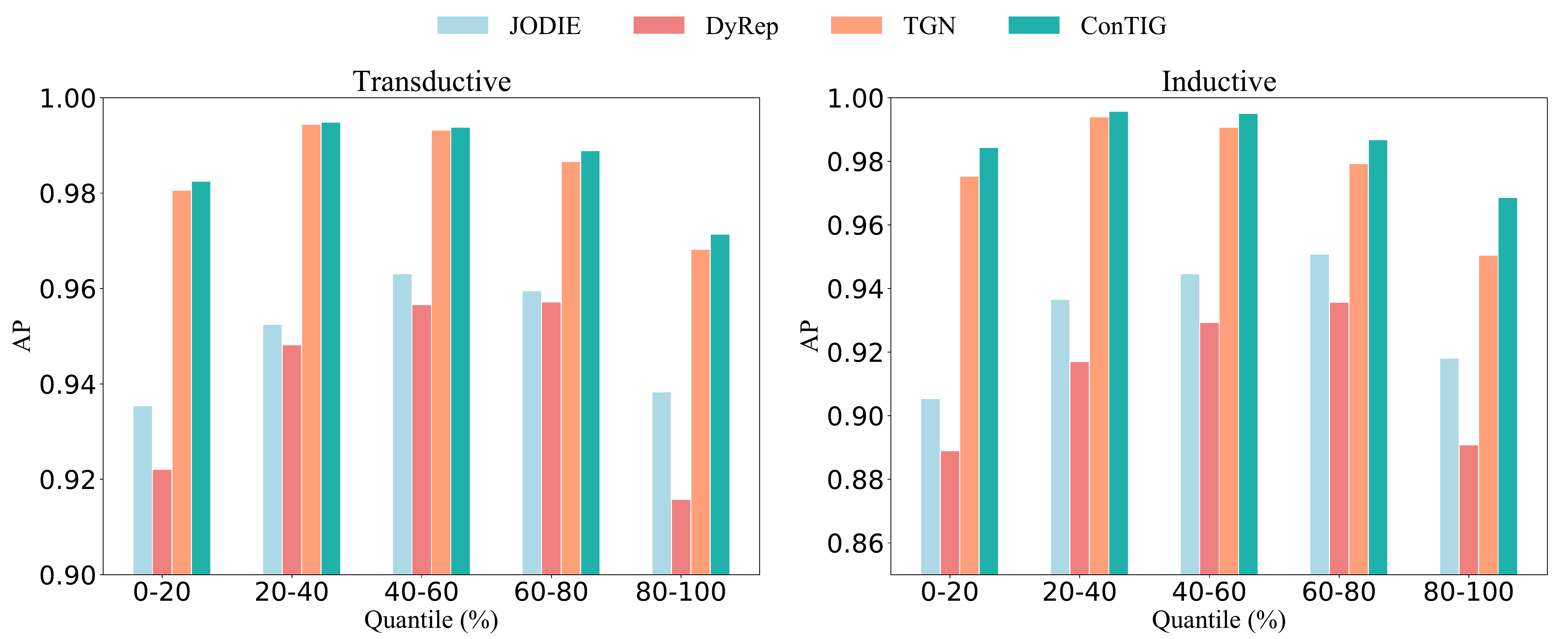}
\label{Fig.4.1}}

\subfloat[CollegeMsg ]{\includegraphics[width=3.5in]{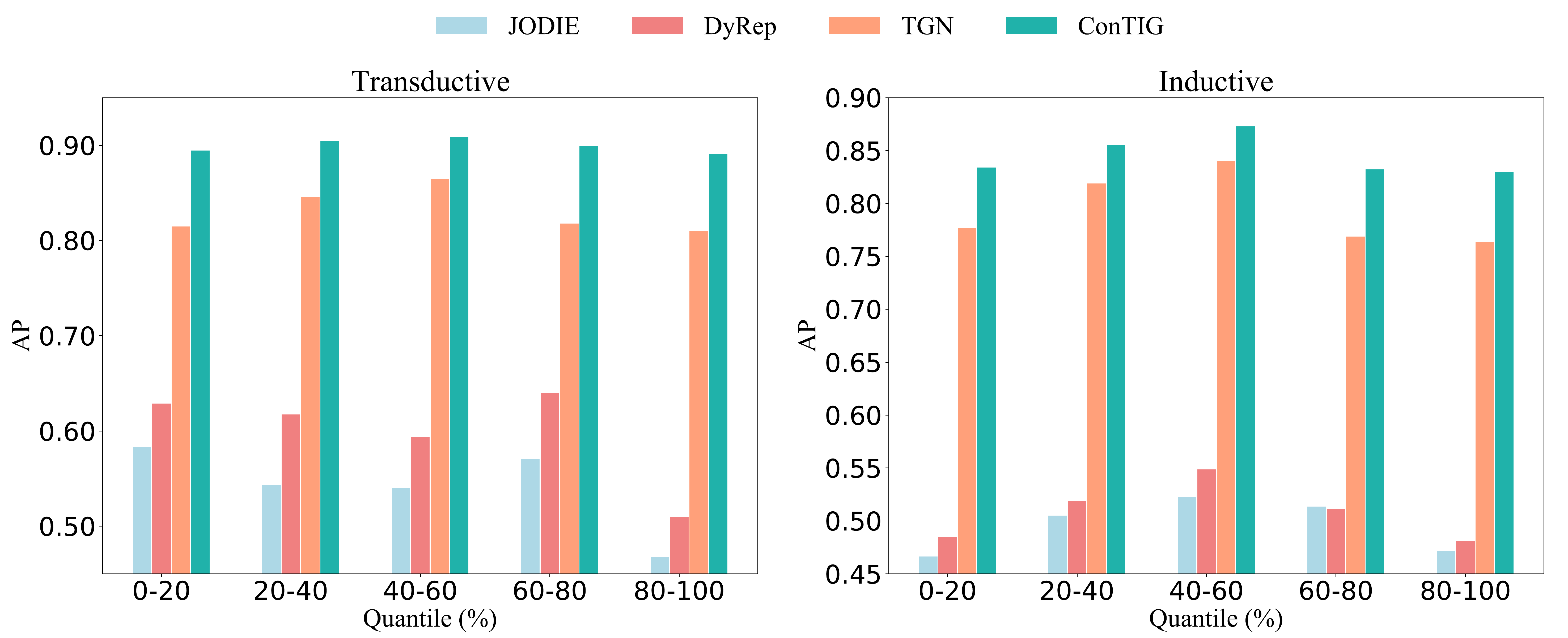}
\label{Fig.4.2}}

\caption{Average Precision(\%) for both the transductive and inductive temporal link prediction on five interaction sets of Wikipedia and CollegeMsg, divided by quantiles of time interval of the interaction.}
\label{Fig.4}
\end{figure}

\begin{table*}
\renewcommand{\arraystretch}{1.3}
\caption{Average Precision(\%) for both the transductive and inductive temporal link prediction on Wikipedia and CollegeMsg. }
\label{tab.4}
\centering
\begin{tabular}{lcccccccc}
    \toprule
    \multirow{2}*{} & \multicolumn{2}{c}{Wikipedia} & \multicolumn{2}{c}{Reddit} &
    \multicolumn{2}{c}{Mooc} &
    \multicolumn{2}{c}{CollegeMsg}\\
    \cline{2-9}
    ~ &  Transductive & Inductive & Transductive & Inductive & Transductive & Inductive  & Transductive & Inductive\\
    \midrule
    ConTIG w/o transform & $95.99$ & $96.76$ & $98.25$ & $97.88$ & $82.33$ & $75.22$ & $87.66$ & $67.65$ \\
    ConTIG w/o update    & $95.19$ & $95.03$ & $95.85$ & $93.36$ & $75.51$ & $73.10$ & $82.78$ & $80.90$ \\
    ConTIG w/o adaptive  & $98.66$ & $98.36$ & $98.75$ & $98.15$ & $68.56$ & $62.42$ & $87.72$ & $84.23$ \\
    ConTIG w/o latest    & $98.50$ & $98.45$ & $98.45$ & $98.79$ & $90.25$ & $86.76$ & $90.66$ & $85.42$ \\
    ConTIG w/o neighbor  & $98.38$ & $98.25$ & $98.02$ & $96.95$ & $84.03$ & $79.57$ & $89.08$ & $83.13$ \\
    ConTIG w/o inherent  & $\textbf{98.75}$ & $98.56$ & $98.79$ & $98.35$ & $86.74$ & $84.27$ & $90.66$ & $82.72$ \\
    ConTIG               & $98.61$ &   $\textbf{98.58}$ & $\textbf{98.79}$ & $\textbf{98.43}$ & $\textbf{90.56}$ & $\textbf{87.31}$ & $\textbf{91.40}$ & $\textbf{86.42}$ \\
    \bottomrule
\end{tabular}
\end{table*}

The goal of this task is to observe the link prediction performance of our method for interaction sets with different time intervals. To categorize the interactions, first, we calculate the time interval $\Delta t$ between each interaction and its latest interaction as mentioned in Section \ref{Sec 3.1}, and describe the distribution of the time intervals of interactions. Then, we equally divide the interactions into five sets according to the four quantiles: 20\%, 40\%, 60\%, 80\%, where the quantiles are cut points dividing the interactions into continuous intervals with equal probabilities in terms of the intervals $\Delta t$ of interactions. For example, the interactions in 0 - 20\% set are of shorter intervals, while the interactions in 80-100\% set are of longer intervals. Finally, we calculate the AP score for each set.

Our results of comparison between our method and dynamic graph baselines in temporal link prediction task on five interaction sets of Wikipedia and CollegeMsg are shown in Fig. \ref{Fig.4}. Comparing the AP results in each set, we find that our method outperforms TGN, especially in long-interval interactions (e.g., the frequency interval 60-80\% and 80-100\%), demonstrating the superiority of ConTIG in capturing continuous dynamics of node representations. And we argue that the long-interval link prediction is more beneficial to practical applications, because quite a few users are always inactive on social networks, citation networks and other user-item interaction systems.

\subsection{Ablation Study}

To further investigate the effect of each component in our model, we compare ConTIG with its variants as follows: 
\begin{itemize}
    \item ConTIG w/o transform: ConTIG without the transform module (i.e. graph attention layer).
    \item ConTIG w/o update: ConTIG without the update module (i.e. continuous inference block). 
    \item ConTIG w/o adaptive: ConTIG without adaptively fusing three factors in update module. We replace it by directly adding the three factors in the update module.
    \item ConTIG w/o latest: ConTIG without the latest interaction factor in update module.
    \item ConTIG w/o neighbor: ConTIG without the neighbor features factor in update module.
    \item ConTIG w/o inherent: ConTIG without the inherent characteristics factor in update module.
\end{itemize}

Table \ref{tab.4} shows the AP results of each model. ConTIG performs better than ConTIG w/o transform and ConTIG w/o update by a large margin, which demonstrates the effectiveness of the main modules of our model. Especially, the introduction of the continuous update module significantly improves the results, indicating that the latest knowledge between two consecutive interactions is an essential feature to learn the node representations. By further modeling and aggregating the historical interaction information of nodes, ConTIG consistently improves the performance, showing the importance of the historical interaction information.

Specifically, we investigate the effect of each component in update module. By removing the adaptive fusion approach, the performance of ConTIG w/o adaptive degrades obviously, pointing out that the adaptive fusion is a key factor to the success of the ODE solver in update module. It helps the network to focus on the most correlated factor to update the node state, and adaptively fuses three factors in a data-dependent way. Besides, ConTIG outperforms ConTIG w/o latest, ConTIG w/o neighbor and ConTIG w/o inherent by a small margin, which indicates that the three factors is of great importance for learning the change of node state in the update module. In particular, by comparing ConTIG with ConTIG w/o latest, ConTIG w/o neighbor and ConTIG w/o inherent, respectively, we observe that the neighbor features contributes most to the performance in all datasets, which indicates the importance of neighbors in the latest changing of node states. Simultaneously, for ConTIG w/o inherent, we observe that it has obvious effects on the datasets without edges features (i.e., Mooc and CollegeMsg), although it has tiny effects on the datasets with edge features (i.e., Wikipedia and Reddit).

\subsection{Parameter Sensitivity} \label{Sec 4.6}

\begin{figure}[!t]
\centering

\subfloat[end time in ODE $\Delta t$]{\includegraphics[width=3.5in]{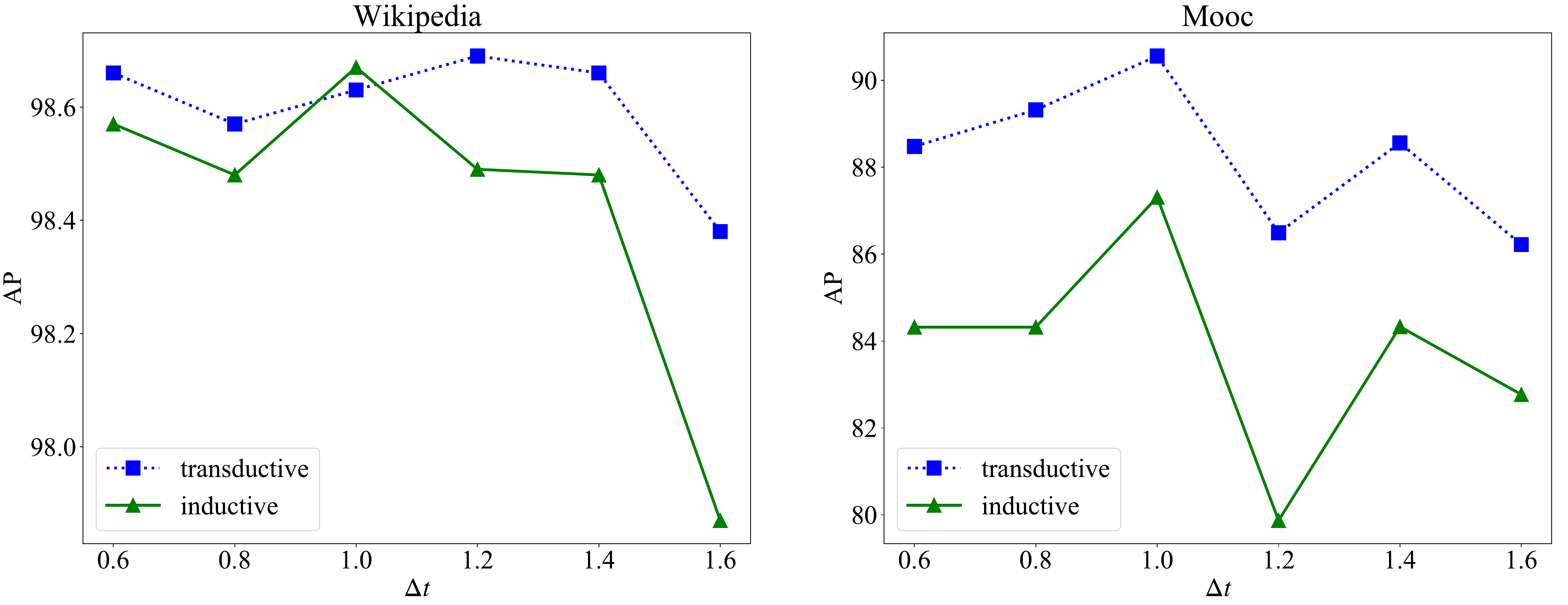}
\label{Fig.5.1}}

\subfloat[number of neighbors $k$]{\includegraphics[width=3.5in]{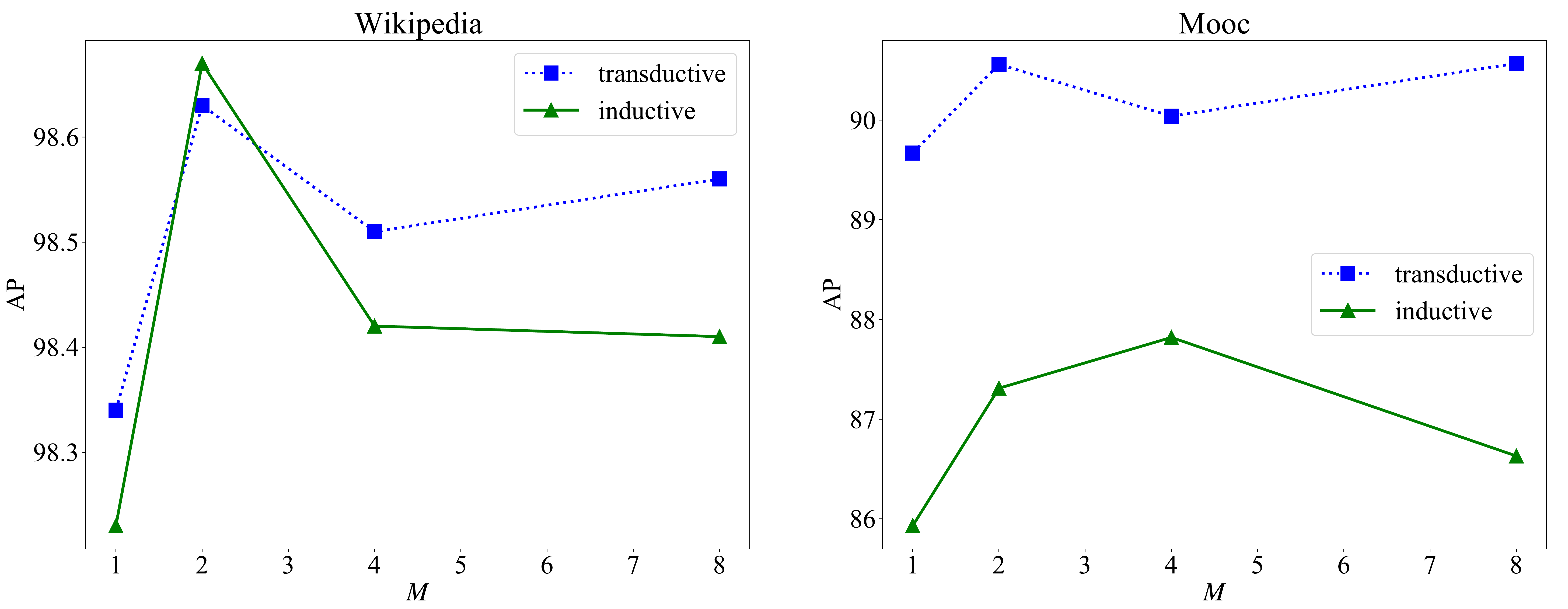}
\label{Fig.5.2}}

\subfloat[number of heads $M$]{\includegraphics[width=3.5in]{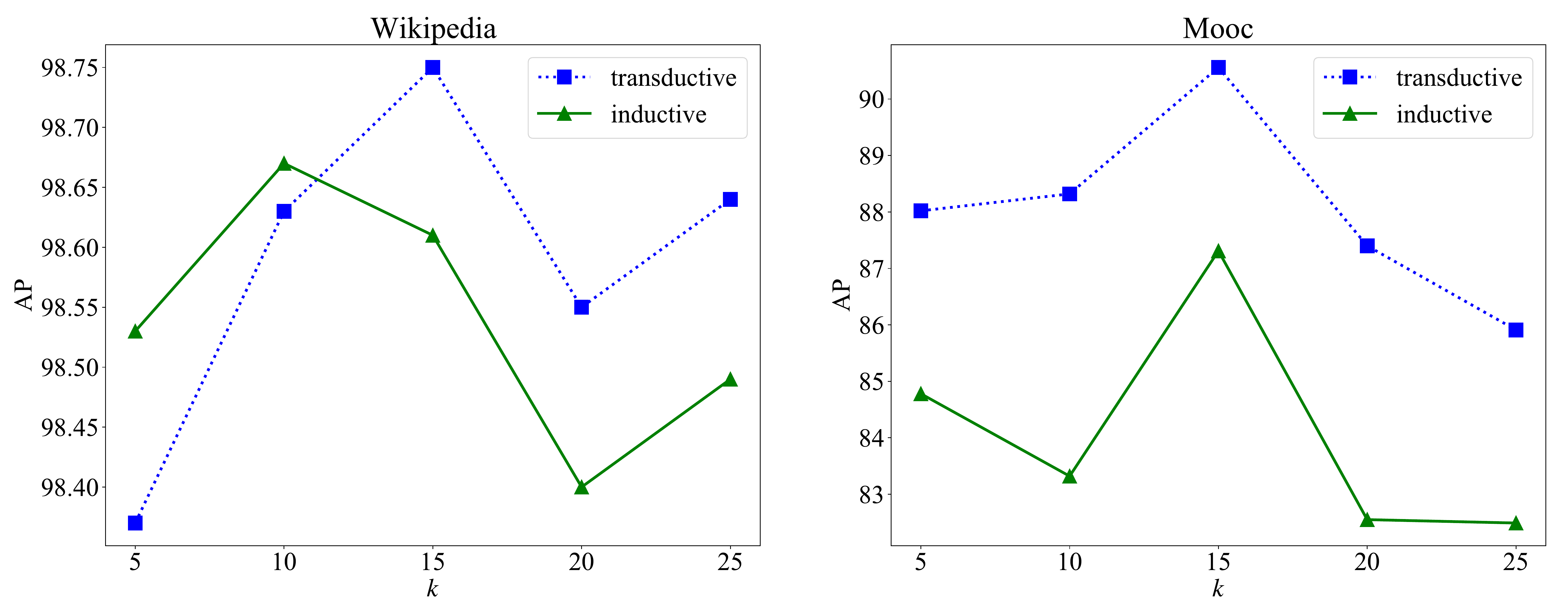}
\label{Fig.5.3}}

\subfloat[node embedding dimension $d$]{\includegraphics[width=3.5in]{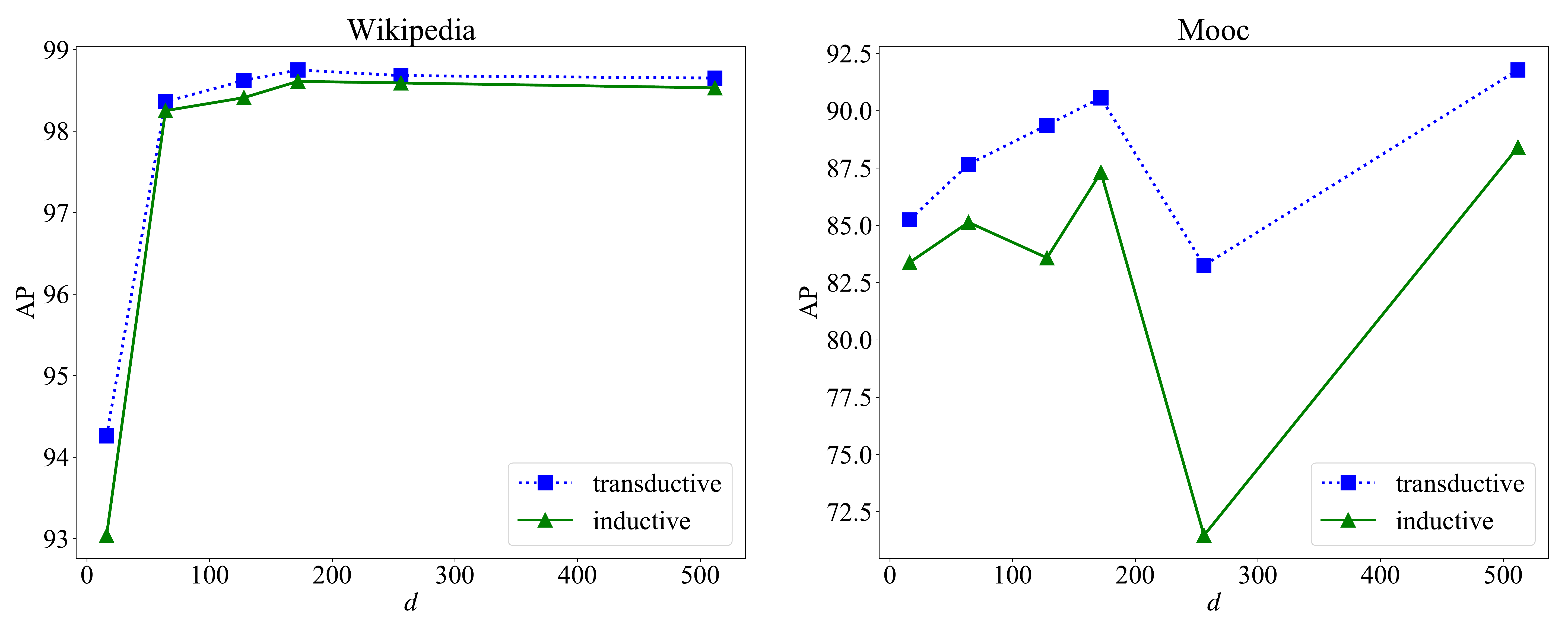}
\label{Fig.5.4}}

\subfloat[time embedding dimension $d^T$]{\includegraphics[width=3.5in]{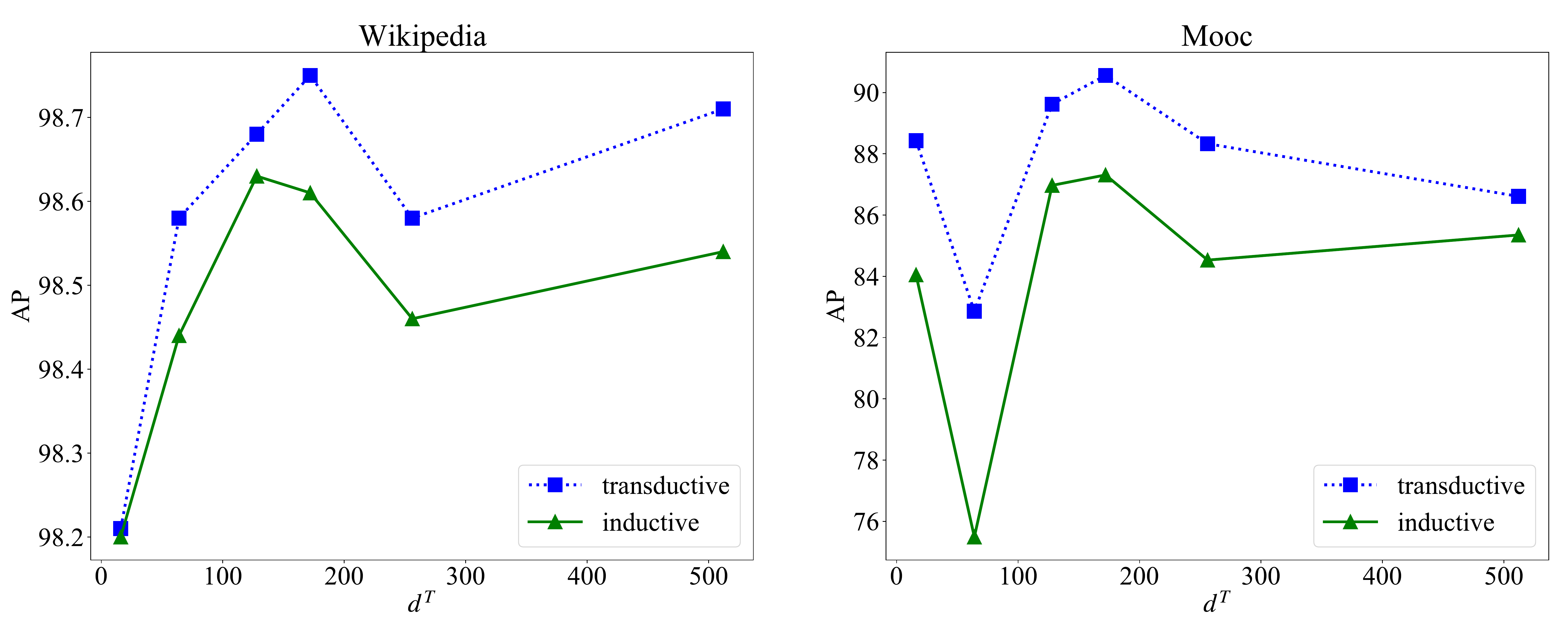}
\label{Fig.5.5}}

\caption{Average Precision(\%) for both transductive and inductive temporal link prediction on Wikipedia and Mooc with different hyper-parameters.}
\label{Fig.5}
\end{figure}

\begin{figure}[!t]
\centering 
\includegraphics[width=0.5\textwidth]{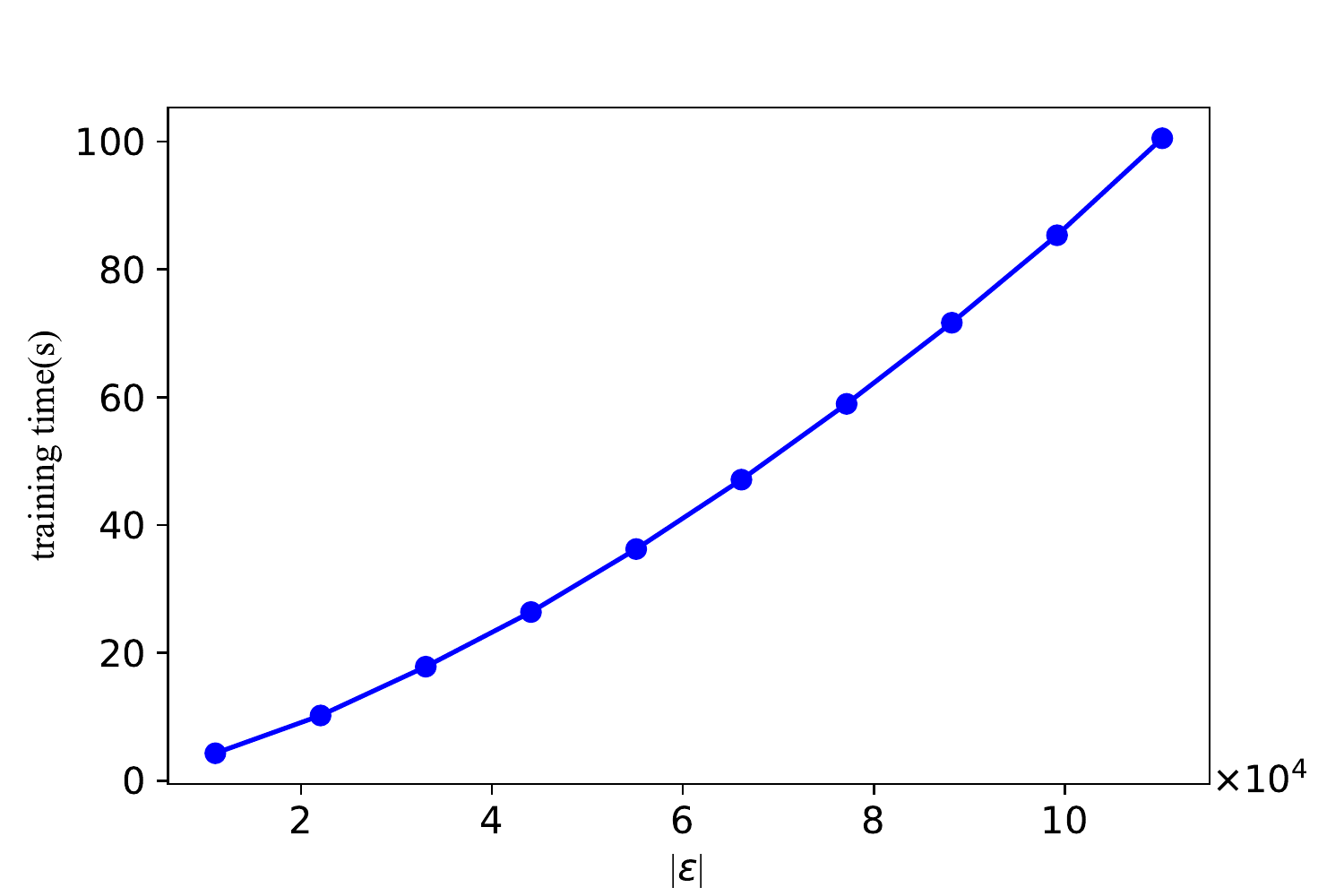}
\caption{Mean time over one epoch during the entire ConTIG training on Wikipedia with different $\left |\mathcal{E}  \right |$ for training.} 
\label{Fig.6}
\end{figure}

There are several hyper-parameters in our proposed method. It is necessary to analyze how these parameters influence the performance of our model in temporal link prediction task on the above datasets. In detail, these parameters include the end time $\Delta t$ in continuous inference block, the number of neighbors $k$, heads $M$ in graph attention layer, node embedding dimension $d$ and time embedding dimension $d^T$. We choose different values for them and use the AP score to evaluate their performance.

\textbf{End time in continuous inference block} (See Section \ref{sec 4.3}). An important parameter is how long used to update the node states in the continuous inference block. In our experiment the $\Delta t$ ranges from 0.6 to 1.6 and the node and time embedding dimension is fixed to 172. As shown in Fig.\ref{Fig.5.1}, as $\Delta t$ is larger, the AP score first increases and then decreases, demonstrating that more time for learning node changes between two consecutive interactions could yield better performance. However, when the updating time is very long, it would make the current node over rely on the latest information, and forget the historical information, which hinders the performance. 

\textbf{Number of neighbors} (See Section \ref{sec 4.4}). The influence of the number of neighbors will then be evaluated, which control the amount of historical interactive information is considered in the transform module. As shown in Fig.\ref{Fig.5.2}, in general, as the number of neighborhood $k$ becomes larger, the model could achieve better performance because more historical information is considered. However, when the number of neighborhood is very large, it would introduce useless or overdue information into the learning process and thus degrade the performance.

\textbf{Heads in graph attention layer} (See Section \ref{sec.4.4}).  Fig.\ref{Fig.5.3} show the AP results under different number of heads $M$ in the multi-head attention. We observed that the AP score first increases and then decreases or stabilize, and there is a relatively stable and good performance at two-head attention. It means that less-head attention will make the results unstable and offset, but attention with many head may pay attention to useless or unrelated information.

\textbf{Node embedding dimension.} The influence of different node embedding dimensions 16; 64; 128; 172; 256 on our model is shown in Fig. \ref{Fig.5.4}. We discover the same trend on two datasets that as the dimension rises from a small value, the performance of ConTIG will improve rapidly and it becomes relative stable while the size is large. The potential reason is that a higher dimension enables the model to learn more information in the latent space. However, it would make computational consumption become very large, so we choose a reasonable dimension with the best performance (i.e., 172).

\textbf{Time embedding dimension.} The influence of different time embedding dimensions 16; 64; 128; 172; 256 on our model is shown in Fig. \ref{Fig.5.5}. We observe the similar results with node embedding dimension, as $d^T$ is larger, the AP score first increases and then tends to be stable. Different from node embedding dimension, there is a fluctuation in the high-dimension time embedding on Wikipedia and low-dimension time embedding on Mooc, which means high dimension and low dimensions may both hinder the performance, and shows the importance of choosing a reasonable time embedding dimension (i.e., 172).





\subsection{Complexity Evaluation}

Here, we examine how the training time of ConTIG depends on the number of edges $\left |\mathcal{E}  \right |$ which are used for training. We record the runtimes of ConTIG for training one epoch on the Wikipedia dataset using the best parameters in Section \ref{Sec 4.6}. Fig.\ref{Fig.6} shows that the entire runtime for one-epoch training is close to linear with $\left |\mathcal{E}  \right |$. This evaluation demonstrates our method’s advantage of being scalable to process long edge streams.

\section{Conclusion}
In this paper, we present ConTIG, a novel representation learning method to capture the continuous dynamic of node embedding trajectories by identifying three-fold factors (i.e., latest interaction, neighbor features and inherent characteristics). ConTIG contains two modules: a update module to learn the node embedding trajectories and a transform module to generate the future representations according to the historical interaction information. Experiments results demonstrate that ConTIG achieves state-of-the-art performances, especially for long-interval interactions on temporal link prediction tasks. In the future, besides node state trajectory, we plan to pay more attention to community evolution, exploring the impact of community on individuals during the graph evolution.



\ifCLASSOPTIONcompsoc
  \section*{Acknowledgments}
\else
  \section*{Acknowledgment}
\fi

The research is supported by Natural Science Foundation of China (61872306), Xiamen Science and Technology Bureau (3502Z20193017) and Fundamental Research Funds for the Central Universities (20720200031).

\ifCLASSOPTIONcaptionsoff
  \newpage
\fi



%




\bibliographystyle{IEEEtran}
\bibliography{references}

\begin{thebibliography}{10}
\providecommand{\url}[1]{#1}
\csname url@samestyle\endcsname
\providecommand{\newblock}{\relax}
\providecommand{\bibinfo}[2]{#2}
\providecommand{\BIBentrySTDinterwordspacing}{\spaceskip=0pt\relax}
\providecommand{\BIBentryALTinterwordstretchfactor}{4}
\providecommand{\BIBentryALTinterwordspacing}{\spaceskip=\fontdimen2\font plus
\BIBentryALTinterwordstretchfactor\fontdimen3\font minus
  \fontdimen4\font\relax}
\providecommand{\BIBforeignlanguage}[2]{{%
\expandafter\ifx\csname l@#1\endcsname\relax
\typeout{** WARNING: IEEEtran.bst: No hyphenation pattern has been}%
\typeout{** loaded for the language `#1'. Using the pattern for}%
\typeout{** the default language instead.}%
\else
\language=\csname l@#1\endcsname
\fi
#2}}
\providecommand{\BIBdecl}{\relax}
\BIBdecl

\bibitem{survey_grl_arxiv_2017}
W.~L. Hamilton, R.~Ying, and J.~Leskovec, ``Representation learning on graphs:
  Methods and applications,'' \emph{arXiv preprint arXiv:1709.05584}, 2017.

\bibitem{survey_ne_tkde_2018}
P.~Cui, X.~Wang, J.~Pei, and W.~Zhu, ``A survey on network embedding,''
  \emph{IEEE Transactions on Knowledge and Data Engineering}, vol.~31, no.~5,
  pp. 833--852, 2018.

\bibitem{deepwalk_kdd_2014}
B.~Perozzi, R.~Al-Rfou, and S.~Skiena, ``Deepwalk: Online learning of social
  representations,'' in \emph{Proceedings of the 20th ACM SIGKDD international
  conference on Knowledge discovery and data mining}, 2014, pp. 701--710.

\bibitem{line_www_2015}
J.~Tang, M.~Qu, M.~Wang, M.~Zhang, J.~Yan, and Q.~Mei, ``Line: Large-scale
  information network embedding,'' in \emph{Proceedings of the 24th
  international conference on world wide web}, 2015, pp. 1067--1077.

\bibitem{node2vec_kdd_2016}
A.~Grover and J.~Leskovec, ``node2vec: Scalable feature learning for
  networks,'' in \emph{Proceedings of the 22nd ACM SIGKDD international
  conference on Knowledge discovery and data mining}, 2016, pp. 855--864.

\bibitem{survey_tnnls_2020}
Z.~Wu, S.~Pan, F.~Chen, G.~Long, C.~Zhang, and S.~Y. Philip, ``A comprehensive
  survey on graph neural networks,'' \emph{IEEE transactions on neural networks
  and learning systems}, vol.~32, no.~1, pp. 4--24, 2020.

\bibitem{gcn_iclr_2017}
\BIBentryALTinterwordspacing
T.~N. Kipf and M.~Welling, ``Semi-supervised classification with graph
  convolutional networks,'' in \emph{5th International Conference on Learning
  Representations, {ICLR} 2017, Toulon, France, April 24-26, 2017, Conference
  Track Proceedings}.\hskip 1em plus 0.5em minus 0.4em\relax OpenReview.net,
  2017. [Online]. Available: \url{https://openreview.net/forum?id=SJU4ayYgl}
\BIBentrySTDinterwordspacing

\bibitem{graphsage_nips_2017}
\BIBentryALTinterwordspacing
W.~L. Hamilton, Z.~Ying, and J.~Leskovec, ``Inductive representation learning
  on large graphs,'' in \emph{Advances in Neural Information Processing Systems
  30: Annual Conference on Neural Information Processing Systems 2017, 4-9
  December 2017, Long Beach, CA, {USA}}, I.~Guyon, U.~von Luxburg, S.~Bengio,
  H.~M. Wallach, R.~Fergus, S.~V.~N. Vishwanathan, and R.~Garnett, Eds., 2017,
  pp. 1024--1034. [Online]. Available:
  \url{http://papers.nips.cc/paper/6703-inductive-representation-learning-on-large-graphs}
\BIBentrySTDinterwordspacing

\bibitem{gat_iclr_2018}
\BIBentryALTinterwordspacing
P.~Velickovic, G.~Cucurull, A.~Casanova, A.~Romero, P.~Li{\`{o}}, and
  Y.~Bengio, ``Graph attention networks,'' in \emph{6th International
  Conference on Learning Representations, {ICLR} 2018, Vancouver, BC, Canada,
  April 30 - May 3, 2018, Conference Track Proceedings}.\hskip 1em plus 0.5em
  minus 0.4em\relax OpenReview.net, 2018. [Online]. Available:
  \url{https://openreview.net/forum?id=rJXMpikCZ}
\BIBentrySTDinterwordspacing

\bibitem{survey_2020_jmlr}
S.~M. Kazemi, R.~Goel, K.~Jain, I.~Kobyzev, A.~Sethi, P.~Forsyth, and
  P.~Poupart, ``Representation learning for dynamic graphs: A survey.''
  \emph{Journal of Machine Learning Research}, vol.~21, no.~70, pp. 1--73,
  2020.

\bibitem{survey_2020_arxiv}
J.~Skarding, B.~Gabrys, and K.~Musial, ``Foundations and modelling of dynamic
  networks using dynamic graph neural networks: A survey,'' \emph{arXiv
  preprint arXiv:2005.07496}, 2020.

\bibitem{survey_arxiv_2021}
G.~Xue, M.~Zhong, J.~Li, J.~Chen, C.~Zhai, and R.~Kong, ``Dynamic network
  embedding survey,'' \emph{arXiv preprint arXiv:2103.15447}, 2021.

\bibitem{burstgraph_2019_ijcai}
\BIBentryALTinterwordspacing
Y.~Zhao, X.~Wang, H.~Yang, L.~Song, and J.~Tang, ``Large scale evolving graphs
  with burst detection,'' in \emph{Proceedings of the Twenty-Eighth
  International Joint Conference on Artificial Intelligence, {IJCAI} 2019,
  Macao, China, August 10-16, 2019}, S.~Kraus, Ed.\hskip 1em plus 0.5em minus
  0.4em\relax ijcai.org, 2019, pp. 4412--4418. [Online]. Available:
  \url{https://doi.org/10.24963/ijcai.2019/613}
\BIBentrySTDinterwordspacing

\bibitem{evolvegcn_2020_aaai}
A.~Pareja, G.~Domeniconi, J.~Chen, T.~Ma, T.~Suzumura, H.~Kanezashi, T.~Kaler,
  T.~B. Schardl, and C.~E. Leiserson, ``Evolvegcn: Evolving graph convolutional
  networks for dynamic graphs.'' in \emph{AAAI}, 2020, pp. 5363--5370.

\bibitem{grade_2020_arxiv}
S.~Spasov, A.~Di~Stefano, P.~Li{\`o}, and J.~Tang, ``Grade: Graph dynamic
  embedding,'' \emph{arXiv preprint arXiv:2007.08060}, 2020.

\bibitem{ctgcn_tkde_2020}
J.~Liu, C.~Xu, C.~Yin, W.~Wu, and Y.~Song, ``K-core based temporal graph
  convolutional network for dynamic graphs,'' \emph{IEEE Transactions on
  Knowledge and Data Engineering}, 2020.

\bibitem{dtine_tkde_2020}
M.~Gong, S.~Ji, Y.~Xie, Y.~Gao, and A.~Qin, ``Exploring temporal information
  for dynamic network embedding,'' \emph{IEEE Transactions on Knowledge and
  Data Engineering}, 2020.

\bibitem{jiao2021temporal}
P.~Jiao, X.~Guo, X.~Jing, D.~He, H.~Wu, S.~Pan, M.~Gong, and W.~Wang,
  ``Temporal network embedding for link prediction via vae joint attention
  mechanism,'' \emph{IEEE Transactions on Neural Networks and Learning
  Systems}, 2021.

\bibitem{tdgnn_www_2020}
L.~Qu, H.~Zhu, Q.~Duan, and Y.~Shi, ``Continuous-time link prediction via
  temporal dependent graph neural network,'' in \emph{Proceedings of The Web
  Conference 2020}, 2020, pp. 3026--3032.

\bibitem{gtea_2020_arxiv}
Y.~Li, D.~S.~H. Tam, S.~Xie, X.~Liu, Q.~F. Ying, W.~C. Lau, D.~M. Chiu, and
  S.~Z. Chen, ``Gtea: Representation learning for temporal interaction graphs
  via edge aggregation,'' \emph{arXiv preprint arXiv:2009.05266}, 2020.

\bibitem{ctdne_www_2018}
G.~H. Nguyen, J.~B. Lee, R.~A. Rossi, N.~K. Ahmed, E.~Koh, and S.~Kim,
  ``Continuous-time dynamic network embeddings,'' in \emph{Companion
  Proceedings of the The Web Conference 2018}, 2018, pp. 969--976.

\bibitem{ties_2020_kdd}
N.~Noorshams, S.~Verma, and A.~Hofleitner, ``Ties: Temporal interaction
  embeddings for enhancing social media integrity at facebook,'' in
  \emph{Proceedings of the 26th ACM SIGKDD International Conference on
  Knowledge Discovery \& Data Mining}, 2020, pp. 3128--3135.

\bibitem{tgat_iclr_2020}
\BIBentryALTinterwordspacing
D.~Xu, C.~Ruan, E.~K{\"{o}}rpeoglu, S.~Kumar, and K.~Achan, ``Inductive
  representation learning on temporal graphs,'' in \emph{8th International
  Conference on Learning Representations, {ICLR} 2020, Addis Ababa, Ethiopia,
  April 26-30, 2020}.\hskip 1em plus 0.5em minus 0.4em\relax OpenReview.net,
  2020. [Online]. Available: \url{https://openreview.net/forum?id=rJeW1yHYwH}
\BIBentrySTDinterwordspacing

\bibitem{tgn_icmlgrl_2020}
\BIBentryALTinterwordspacing
E.~Rossi, B.~Chamberlain, F.~Frasca, D.~Eynard, F.~Monti, and M.~M. Bronstein,
  ``Temporal graph networks for deep learning on dynamic graphs,'' \emph{CoRR},
  vol. abs/2006.10637, 2020. [Online]. Available:
  \url{https://arxiv.org/abs/2006.10637}
\BIBentrySTDinterwordspacing

\bibitem{jodie_kdd_2019}
S.~Kumar, X.~Zhang, and J.~Leskovec, ``Predicting dynamic embedding trajectory
  in temporal interaction networks,'' in \emph{Proceedings of the 25th {ACM}
  {SIGKDD} International Conference on Knowledge Discovery {\&} Data Mining,
  {KDD} 2019, Anchorage, AK, USA, August 4-8, 2019}, A.~Teredesai, V.~Kumar,
  Y.~Li, R.~Rosales, E.~Terzi, and G.~Karypis, Eds., 2019, pp. 1269--1278.

\bibitem{tigecmn_www_2020}
Z.~Zhang, J.~Bu, M.~Ester, J.~Zhang, C.~Yao, Z.~Li, and C.~Wang, ``Learning
  temporal interaction graph embedding via coupled memory networks,'' in
  \emph{Proceedings of The Web Conference 2020}, 2020, pp. 3049--3055.

\bibitem{discrete_dynamic_iclr_2020}
\BIBentryALTinterwordspacing
K.~Oono and T.~Suzuki, ``Graph neural networks exponentially lose expressive
  power for node classification,'' in \emph{8th International Conference on
  Learning Representations, {ICLR} 2020, Addis Ababa, Ethiopia, April 26-30,
  2020}.\hskip 1em plus 0.5em minus 0.4em\relax OpenReview.net, 2020. [Online].
  Available: \url{https://openreview.net/forum?id=S1ldO2EFPr}
\BIBentrySTDinterwordspacing

\bibitem{cgnn_icml_2020}
\BIBentryALTinterwordspacing
L.~A.~C. Xhonneux, M.~Qu, and J.~Tang, ``Continuous graph neural networks,'' in
  \emph{Proceedings of the 37th International Conference on Machine Learning,
  {ICML} 2020, 13-18 July 2020, Virtual Event}, ser. Proceedings of Machine
  Learning Research, vol. 119.\hskip 1em plus 0.5em minus 0.4em\relax {PMLR},
  2020, pp. 10\,432--10\,441. [Online]. Available:
  \url{http://proceedings.mlr.press/v119/xhonneux20a.html}
\BIBentrySTDinterwordspacing

\bibitem{neuralode_nips_2018}
R.~T. Chen, Y.~Rubanova, J.~Bettencourt, and D.~K. Duvenaud, ``Neural ordinary
  differential equations,'' in \emph{Advances in neural information processing
  systems}, 2018, pp. 6571--6583.

\bibitem{le_nips_2001}
M.~Belkin and P.~Niyogi, ``Laplacian eigenmaps and spectral techniques for
  embedding and clustering.'' in \emph{Nips}, vol.~14, no.~14, 2001, pp.
  585--591.

\bibitem{gf_www_2013}
A.~Ahmed, N.~Shervashidze, S.~Narayanamurthy, V.~Josifovski, and A.~J. Smola,
  ``Distributed large-scale natural graph factorization,'' in \emph{Proceedings
  of the 22nd international conference on World Wide Web}, 2013, pp. 37--48.

\bibitem{pan2019learning}
S.~Pan, R.~Hu, S.-f. Fung, G.~Long, J.~Jiang, and C.~Zhang, ``Learning graph
  embedding with adversarial training methods,'' \emph{IEEE transactions on
  cybernetics}, vol.~50, no.~6, pp. 2475--2487, 2020.

\bibitem{attention_nips_2017}
A.~Vaswani, N.~Shazeer, N.~Parmar, J.~Uszkoreit, L.~Jones, A.~N. Gomez,
  {\L}.~Kaiser, and I.~Polosukhin, ``Attention is all you need,'' in
  \emph{Advances in neural information processing systems}, 2017, pp.
  5998--6008.

\bibitem{tNodeEmbed_2019_ijcai}
\BIBentryALTinterwordspacing
U.~Singer, I.~Guy, and K.~Radinsky, ``Node embedding over temporal graphs,'' in
  \emph{Proceedings of the Twenty-Eighth International Joint Conference on
  Artificial Intelligence, {IJCAI} 2019, Macao, China, August 10-16, 2019},
  S.~Kraus, Ed.\hskip 1em plus 0.5em minus 0.4em\relax ijcai.org, 2019, pp.
  4605--4612. [Online]. Available:
  \url{https://doi.org/10.24963/ijcai.2019/640}
\BIBentrySTDinterwordspacing

\bibitem{dysat_2020_icdm}
A.~Sankar, Y.~Wu, L.~Gou, W.~Zhang, and H.~Yang, ``Dysat: Deep neural
  representation learning on dynamic graphs via self-attention networks,'' in
  \emph{Proceedings of the 13th International Conference on Web Search and Data
  Mining}, 2020, pp. 519--527.

\bibitem{htgn_kdd_2021}
M.~Yang, M.~Zhou, M.~Kalander, Z.~Huang, and I.~King, ``Discrete-time temporal
  network embedding via implicit hierarchical learning in hyperbolic space,''
  in \emph{Proc. of 2021 ACM SIGKDD Int. Conf. on Knowledge Discovery and Data
  Mining (KDD'21)}, 2021.

\bibitem{dynamicppe_kdd_2021}
X.~Guo, B.~Zhou, and S.~Skiena, ``Subset node representation learning over
  large dynamic graphs,'' in \emph{Proc. of 2021 ACM SIGKDD Int. Conf. on
  Knowledge Discovery and Data Mining (KDD'21)}, 2021.

\bibitem{dyngem_2018_arxiv}
P.~Goyal, N.~Kamra, X.~He, and Y.~Liu, ``Dyngem: Deep embedding method for
  dynamic graphs,'' \emph{arXiv preprint arXiv:1805.11273}, 2018.

\bibitem{trrn_aaai_2021}
D.~Xu, J.~Liang, W.~Cheng, H.~Wei, H.~Chen, and X.~Zhang, ``Transformer-style
  relational reasoning with dynamic memory updating for temporal network
  modeling,'' in \emph{Proceedings of the AAAI Conference on Artificial
  Intelligence}, vol.~35, no.~5, 2021, pp. 4546--4554.

\bibitem{dyrep_2019_iclr}
R.~Trivedi, M.~Farajtabar, P.~Biswal, and H.~Zha, ``Dyrep: Learning
  representations over dynamic graphs,'' in \emph{International Conference on
  Learning Representations}, 2019.

\bibitem{m2dne_cikm_2019}
Y.~Lu, X.~Wang, C.~Shi, P.~S. Yu, and Y.~Ye, ``Temporal network embedding with
  micro-and macro-dynamics,'' in \emph{Proceedings of the 28th ACM
  international conference on information and knowledge management}, 2019, pp.
  469--478.

\bibitem{tcl_arxiv_2021}
L.~Wang, X.~Chang, S.~Li, Y.~Chu, H.~Li, W.~Zhang, X.~He, L.~Song, J.~Zhou, and
  H.~Yang, ``Tcl: Transformer-based dynamic graph modelling via contrastive
  learning,'' \emph{arXiv preprint arXiv:2105.07944}, 2021.

\bibitem{taggen_kdd_2020}
D.~Zhou, L.~Zheng, J.~Han, and J.~He, ``A data-driven graph generative model
  for temporal interaction networks,'' in \emph{Proceedings of the 26th ACM
  SIGKDD International Conference on Knowledge Discovery \& Data Mining}, 2020,
  pp. 401--411.

\bibitem{ad_arxiv_2021}
M.~Liu, Z.~Tu, X.~Xu, and Z.~Wang, ``Learning representation over dynamic graph
  using aggregation-diffusion mechanism,'' \emph{arXiv preprint
  arXiv:2106.01678}, 2021.

\bibitem{motifs_wsdm_2017}
A.~Paranjape, A.~R. Benson, and J.~Leskovec, ``Motifs in temporal networks,''
  in \emph{Proceedings of the tenth ACM international conference on web search
  and data mining}, 2017, pp. 601--610.

\bibitem{caw_iclr_2021}
Y.~Wang, Y.-Y. Chang, Y.~Liu, J.~Leskovec, and P.~Li, ``Inductive
  representation learning in temporal networks via causal anonymous walks,''
  \emph{arXiv preprint arXiv:2101.05974}, 2021.

\bibitem{lmega_kdd_2020}
D.~Fu, D.~Zhou, and J.~He, ``Local motif clustering on time-evolving graphs,''
  in \emph{Proceedings of the 26th ACM SIGKDD International conference on
  knowledge discovery \& data mining}, 2020, pp. 390--400.

\bibitem{ndcn_kdd_2020}
\BIBentryALTinterwordspacing
C.~Zang and F.~Wang, ``Neural dynamics on complex networks,'' in \emph{{KDD}
  '20: The 26th {ACM} {SIGKDD} Conference on Knowledge Discovery and Data
  Mining, Virtual Event, CA, USA, August 23-27, 2020}, R.~Gupta, Y.~Liu,
  J.~Tang, and B.~A. Prakash, Eds.\hskip 1em plus 0.5em minus 0.4em\relax
  {ACM}, 2020, pp. 892--902. [Online]. Available:
  \url{https://doi.org/10.1145/3394486.3403132}
\BIBentrySTDinterwordspacing

\bibitem{tkgode_arxiv_2021}
Z.~Ding, Z.~Han, Y.~Ma, and V.~Tresp, ``Temporal knowledge graph forecasting
  with neural ode,'' \emph{arXiv preprint arXiv:2101.05151}, 2021.

\bibitem{cgode_kdd_2021}
Z.~Huang, Y.~Sun, and W.~Wang, ``Coupled graph ode for learning interacting
  system dynamics,'' in \emph{Proc. of 2021 ACM SIGKDD Int. Conf. on Knowledge
  Discovery and Data Mining (KDD'21)}, 2021.

\bibitem{ftrl_nips_2019}
\BIBentryALTinterwordspacing
D.~Xu, C.~Ruan, E.~K{\"{o}}rpeoglu, S.~Kumar, and K.~Achan, ``Self-attention
  with functional time representation learning,'' in \emph{Advances in Neural
  Information Processing Systems 32: Annual Conference on Neural Information
  Processing Systems 2019, NeurIPS 2019, December 8-14, 2019, Vancouver, BC,
  Canada}, H.~M. Wallach, H.~Larochelle, A.~Beygelzimer,
  F.~d'Alch{\'{e}}{-}Buc, E.~B. Fox, and R.~Garnett, Eds., 2019, pp.
  15\,889--15\,899. [Online]. Available:
  \url{http://papers.nips.cc/paper/9720-self-attention-with-functional-time-representation-learning}
\BIBentrySTDinterwordspacing

\bibitem{gae_arxiv_2016}
\BIBentryALTinterwordspacing
T.~N. Kipf and M.~Welling, ``Variational graph auto-encoders,'' \emph{CoRR},
  vol. abs/1611.07308, 2016. [Online]. Available:
  \url{http://arxiv.org/abs/1611.07308}
\BIBentrySTDinterwordspacing

\end{thebibliography}

\end{document}